\documentclass[useAMS,usenatbib]{mn2e}

\usepackage{graphicx}  
\usepackage{ amssymb }
\usepackage{journals}
\usepackage[dvips]{color}
\newcommand{\teff}{$T_{\rm eff}$} 
\newcommand{\logg}{$\log g$} 
\newcommand{\kms}{km s$^{-1}$}
\newcommand{\vt}{$\xi_t$} 
\newcommand{\fei}{Fe\,{\sc i}}
\newcommand{\feii}{Fe\,{\sc ii}}
\newcommand{\scii}{Sc\,{\sc ii}}
\newcommand{\oi}{O\,{\sc i}}
\newcommand{\nai}{Na\,{\sc i}}
\newcommand{\mgi}{Mg\,{\sc i}}
\newcommand{\ali}{Al\,{\sc i}}
\newcommand{\sii}{Si\,{\sc i}}
\newcommand{\cai}{Ca\,{\sc i}}
\newcommand{\caii}{Ca\,{\sc ii}}
\newcommand{\tii}{Ti\,{\sc i}}
\newcommand{\tiii}{Ti\,{\sc ii}}
\newcommand{\vi}{V\,{\sc i}}
\newcommand{\mni}{Mn\,{\sc i}}
\newcommand{\coi}{Co\,{\sc i}}
\newcommand{\nii}{Ni\,{\sc i}}
\newcommand{\cui}{Cu\,{\sc i}}
\newcommand{\zni}{Zn\,{\sc i}}
\newcommand{\yii}{Y\,{\sc ii}}
\newcommand{\zri}{Zr\,{\sc i}}
\newcommand{\zrii}{Zr\,{\sc ii}}
\newcommand{\moi}{Mo\,{\sc i}}
\newcommand{\sri}{Sr\,{\sc i}}

\newcommand{\pbi}{Pb\,{\sc i}}
\newcommand{\baii}{Ba\,{\sc ii}}
\newcommand{\laii}{La\,{\sc ii}}
\newcommand{\ceii}{Ce\,{\sc ii}}
\newcommand{\ndii}{Nd\,{\sc ii}}
\newcommand{\prii}{Pr\,{\sc ii}}
\newcommand{\smii}{Sm\,{\sc ii}}
\newcommand{\gdii}{Gd\,{\sc ii}}
\newcommand{\tbii}{Tb\,{\sc ii}}
\newcommand{\dyii}{Dy,{\sc ii}}
\newcommand{\erii}{Er\,{\sc ii}}
\newcommand{\tmii}{Tm\,{\sc ii}}
\newcommand{\ybii}{Yb\,{\sc ii}}
\newcommand{\hfii}{Hf\,{\sc ii}}
\newcommand{\euii}{Eu\,{\sc ii}}
\newcommand{\cri}{Cr\,{\sc i}}
\newcommand{\crii}{Cr\,{\sc ii}}

\newcommand\simgt{\lower.3ex\hbox{\gtsima}}

\newcommand{\strom}{Str{\" o}mgren}
\newcommand{\mch}{$m_{\rm CH}$}
\newcommand{\scn}{$S(3839)$}
\newcommand\msun{M$_{\odot}$}
\newcommand{\hst}{$Hubble~Space~Telescope$} 
\def\rpro{\mbox{$r$-process}}
\def\spro{\mbox{$s$-process}}

\def\msun{$M_{\odot}$}
\def\bd{\mbox{BD$+$17$^\circ$~3248}}
\def\loggf{$\log gf$}
\usepackage{placeins}
\usepackage[modulo]{lineno}

\title[Chemical abundances in M2]{Iron and neutron-capture element abundance
variations in the globular cluster M2 (NGC 7089)\thanks{Based in part on data 
collected at Subaru Telescope, which is operated by the National Astronomical
Observatory of Japan. This paper includes data gathered with the 6.5 meter
Magellan Telescopes located at Las Campanas Observatory, Chile.}} 
\author[D.\ Yong et al.]
{David Yong,$^{1}$\thanks{E-mail: david.yong@anu.edu.au}\thanks{Stromlo Fellow}  
Ian U.\ Roederer,$^2$  
Frank Grundahl,$^3$ 
Gary S.\ Da Costa,$^1$ \newauthor 
Amanda I.\ Karakas,$^1$ 
John E.\ Norris,$^1$  
Wako Aoki,$^4$ 
Cherie K.\ Fishlock,$^1$ \newauthor 
A.\ F.\ Marino,$^1$ 
A.\ P.\ Milone$^1$ and 
Luke J.\ Shingles$^1$. \\ 
$^{1}$Research School of Astronomy and Astrophysics, Australian
National University, Canberra, ACT 2611, Australia\\ 
$^{2}$Department of Astronomy, University of Michigan, 500 Church Street, Ann
Arbor, MI 48109, USA\\ 
$^{3}$Stellar Astrophysics Centre, Department of Physics and 
Astronomy, Aarhus University, Ny Munkegade 120, DK-8000 Aarhus C, Denmark\\
$^{4}$National Astronomical Observatory, Mitaka, Tokyo 181-8588,
Japan\\
}
\begin{document}


\pagerange{\pageref{firstpage}$-$\pageref{lastpage}} \pubyear{2013}

\maketitle

\label{firstpage}

\begin{abstract}
We present CN and CH indices and \caii\ triplet metallicities for 34 giant
stars and chemical abundances for 33 elements in 14 giants in the globular
cluster M2. Assuming the program stars are cluster members, our analysis
reveals ($i$) an extreme variation in CN and CH line strengths, ($ii$) a
metallicity dispersion with a dominant peak at [Fe/H] $\approx$ $-$1.7 and
smaller peaks at $-$1.5 and $-$1.0, ($iii$) star-to-star abundance variations
and correlations for the light elements O, Na, Al and Si and ($iv$) a large
(and possibly bimodal) distribution in the abundances of all elements produced
mainly via the $s$-process in solar system material. Following
\citet{roederer11}, we define two groups of stars, ``$r+s$'' and ``$r$-only'',
and subtract the average abundances of the latter from the former group to
obtain a ``$s$-process residual''. This $s$-process residual is remarkably
similar to that found in M22 and in M4 despite the range in metallicity covered
by these three systems.  With recent studies identifying a double subgiant
branch in M2 and a dispersion in Sr and Ba abundances, our spectroscopic
analysis confirms that this globular cluster has experienced a complex
formation history with similarities to M22, NGC 1851 and $\omega$ Centauri. 
\end{abstract}


\section{INTRODUCTION}

Photometric studies have revealed complex structure in the colour-magnitude
diagrams (CMD) of Galactic globular clusters (e.g., see review by
\citealt{piotto09}). The subgiant branch region is of particular interest
because differences in the luminosity of stars at this evolutionary stage
require distinct ages and/or chemical compositions. Any globular cluster that
exhibits a broadened or split subgiant branch must therefore have experienced a
complex, and likely prolonged, chemical enrichment history when compared to
globular clusters with a single subgiant branch population. 

$\omega$ Centauri and M22 (NGC 6656) are two Galactic globular clusters with
multiple subgiant branches (e.g., \citealt{bedin04}; \citealt{marino09}). These
two clusters are also notable for exhibiting a large star-to-star dispersion in
the abundance of Fe-peak and neutron-capture elements (e.g.,
\citealt{norris95,smith00,marino09,marino11,johnson10,roederer11}).  NGC 1851
is another globular cluster with multiple subgiant branches \citep{milone08}.
Although the difference in metallicity between the two populations,
$\Delta$[Fe/H]~$\approx$~0.07~dex \citep{carretta101851}, is less pronounced in
NGC 1851 compared to $\omega$ Cen and M22, a large star-to-star dispersion in
the neutron-capture element abundances is also present (e.g.,
\citealt{yong081851,villanova09,carretta11}). While theoretical studies
indicate that multiple population globular clusters could be formed through
mergers or that some may be the remnants of dwarf galaxies (e.g.,
\citealt{bekki03,carretta10,bekki11,bekki12}), understanding the sequence of
events that produce multiple population globular clusters remains a major
challenge (e.g.,
\citealt{marcolini07,dercole08,dantona10,conroy11,herwig12,vesperini13}). An
important step in advancing our knowledge of the formation of multiple
population globular clusters is to understand the full range of phenomena and
relative frequency present in the Galactic globular cluster system. 

\citet{piotto12} identified five new Galactic globular clusters with broadened
or split subgiant branches based on \hst\ photometry. Their sample included M2
(NGC 7089), a little-studied cluster. \citet{smith90} measured the strengths of
the CN and CH molecular features in a sample of 19 M2 red giants. In addition
to the usual bimodal distribution of CN band strengths \citep{smith87}, they
noted that two objects are CH stars. CH stars are rare in globular clusters,
and at the time of that paper, the only other clusters known to contain CH
stars included the apparently normal cluster M55 as well as the peculiar
systems M22 and $\omega$ Cen. \citet{smolinski11} studied the CN and CH bands
from Sloan Digital Sky Survey spectroscopy in a number of globular clusters
including M2. They did not identify any stars with unusually strong CN or CH in
this cluster, and all of their program stars lie on the canonical red giant
branch (RGB). \citet{lardo12} studied the CN and CH band strengths as well as
the C and N abundances in a sample of 35 M2 red giants.  They also noted the
presence of an additional RGB in the $V$ versus $U-V$ CMD.  Both CH stars
identified by \citet{smith90} are located on the anomalous RGB (see Figure 14
in \citealt{lardo12}).  Examination of the \citet{grundahl99} \strom\
photometry also confirms the peculiar nature of the RGB. While \citet{lardo12}
did not observe any stars on the anomalous RGB, in a subsequent study they
obtained spectra for such stars \citep{lardo13}. Stars belonging to the two
RGBs had distinct C, N, Sr and Ba abundances and \citet{lardo13} argued that M2
has experienced a complex star formation history with similarities to $\omega$
Cen, M22 and NGC 1851. 

High-resolution spectroscopy and chemical abundance measurements for a larger
suite of elements for stars on the canonical and anomalous RGBs of M2 are
essential to reveal the true nature of this multiple population globular
cluster. The purpose of this paper is to measure CN and CH indices and chemical
abundances for a sample of stars in M2 belonging to the canonical and anomalous
RGBs. The sample selection and observations are described in Section 2. Section
3 contains the analysis. The results are presented in Section 4. Section 5
includes a discussion on the nature of this cluster. 

\section{SAMPLE SELECTION AND OBSERVATIONS}

\begin{figure}
\centering
      \includegraphics[width=.95\hsize]{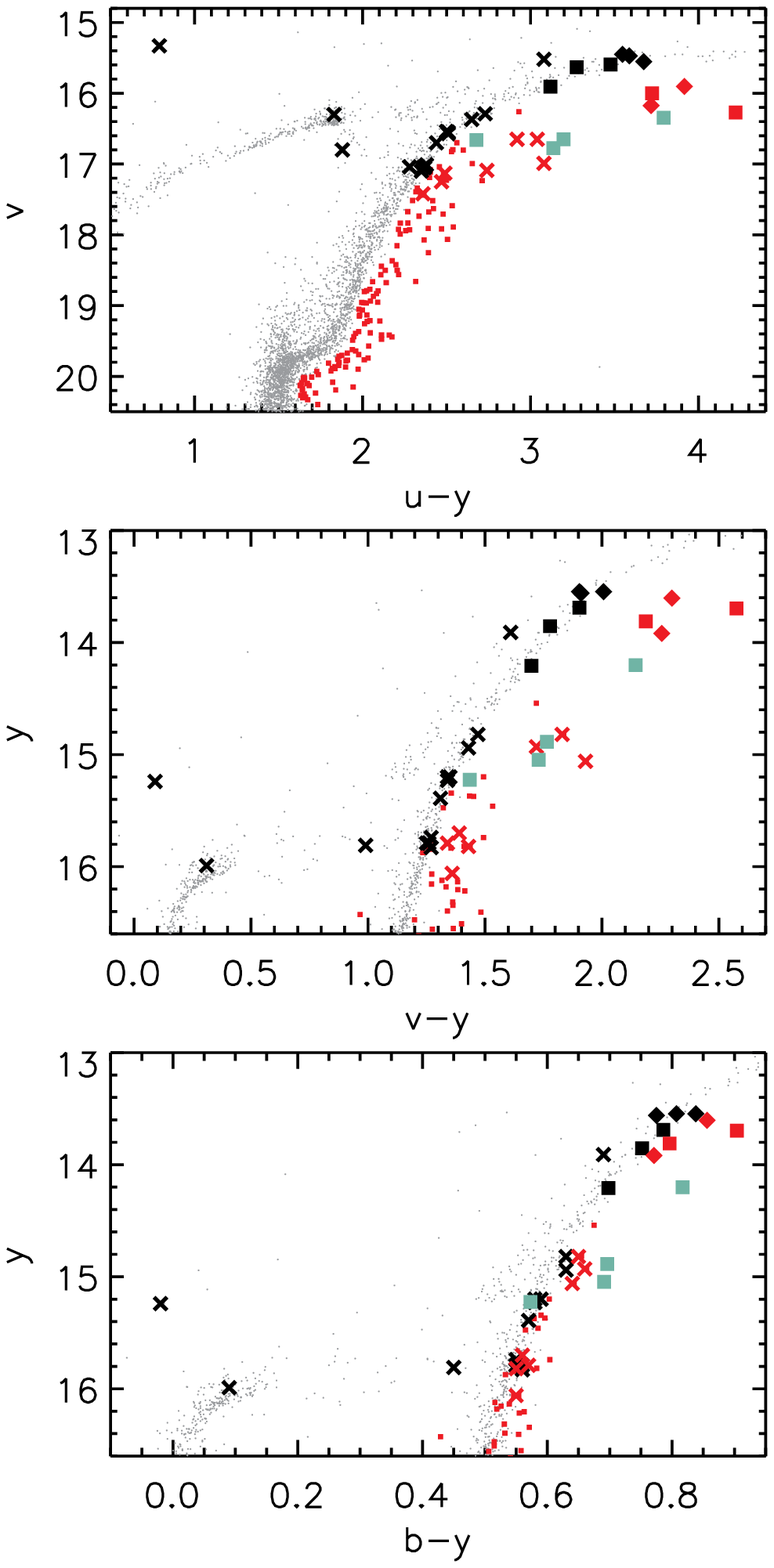}
      \caption{Colour-magnitude diagrams for $v$ versus $u-y$ (upper), $y$
versus $v-y$ (middle) and $y$ versus $b-y$ (lower). The black symbols are stars
that lie on the canonical RGB, AGB or HB as well as one UV bright object. The
red (and aqua) symbols are stars that lie on the anomalous RGB, and were
selected from the $v$ versus $u-y$ CMD, upper panel. (The aqua symbols are the
unusually metal-rich objects as determined from high-resolution spectroscopy.)
Crosses represent stars observed with the AAT. Square symbols (Subaru
Telescope) and diamond symbols (Magellan Telescope) represent objects observed
at high spectral resolution. 
      \label{fig:cmd100} }
\end{figure}

The program stars were selected from the $uvby$ \strom\ photometry by
\citet{grundahl99}. In Figure \ref{fig:cmd100}, we present the $u-y$, $v-y$,
and $b-y$ CMDs. As noted by \citet{lardo12}, we confirm the presence of an
additional RGB sequence. Such stars are highlighted in red and aqua in Figure
\ref{fig:cmd100} and were selected from the $v$ versus $u-y$ CMD (upper panel
in Figure \ref{fig:cmd100}). We refer to these as anomalous RGB stars. (The
reason for using two sets of colours for the anomalous RGB stars will become
clear when we present the chemical abundances for these objects: the red
symbols are stars with [Fe/H] $\approx$ $-$1.5 and the aqua symbols are stars
with [Fe/H] $\approx$ $-$1.0.) While \citet{lardo12} showed that the two CH
stars identified by \citet{smith90} were located on the anomalous giant branch,
neither was included in the \citet{grundahl99} photometry. 

\subsection{Medium-resolution spectroscopic observations} 

We observed candidate M2 members using the AAOmega multi-object spectrograph
\citep{saunders04} on the Anglo-Australian Telescope as part of two separate
observing runs. The two CH stars from \citet{smith90} were also observed. The
first set of observations, obtained on 2010 September 30, used the 1700B blue
and 2000R red gratings.  These provide spectral coverage of 3750\,\AA\/ to
4440\,\AA\/ and 5800\,\AA\/ to 6300\,\AA\/ at resolutions $R$ =
$\lambda$/$\Delta\lambda$ = 3500 and 8000 for the red and blue arms,
respectively.  The second set of observations, from 2011 November 1, employed
the 580V blue and 1700D red gratings. These gratings provide a wavelength
coverage from 3750\,\AA\/ to 5500\,\AA\/ at a resolution of 1300 in the blue
arm, and 8350\,\AA\/ to 8825\,\AA\/ at $R$ = 10000 in the red arm.  In each
case the cluster observations were obtained together with flat-field and arc
lamp calibration exposures.  Data reduction to wavelength-calibrated
sky-subtracted individual stellar spectra was accomplished using the standard
{\sc 2dfdr}\footnote{http://www.aao.gov.au/2df/aaomega/aaomega\_2dfdr.html}
software.  In generating the fibre configurations for the observations, high
priority was given to the stars that lie on the anomalous giant branch. 

Although the same input catalogue was used for both sets of observations, the
combination of different field plates and different available fibre numbers
meant that the two sets of stars observed are not identical.  The first set
contains 22 stars, the second 23 with 11 in common for a total sample of 34
candidate M2 members.  The stars observed are listed in Table \ref{tab:aao}. We
note that 14 stars belong to the anomalous RGB and the two CH stars from
\citet{smith90} may also be regarded as anomalous RGB objects
\citep{lardo12,lardo13}.  Based on their CMD location, we offer some comments
on a handful of the stars observed with AAOmega.   Star NR 82, if a cluster
member, would be classified as a asymptotic giant branch (AGB) star rather
than a RGB star.  Star NR 184, if a cluster member, would be a UV-bright object
lying more than a magnitude above the cluster blue horizontal branch.
Similarly, star NR~648 has the photometric colours and magnitude of a cluster
blue horizontal branch star.  However, the AAOmega spectra (from the second
data set) are that of a cluster-like red giant star.  We have no
straightforward explanation for this anomaly.  Star NR 707, again if it is a
cluster member, has a magnitude and colour that would suggest it is a red
horizontal branch (RHB) star.  M2, however, is not normally considered to have
a RHB population given its dominant blue HB morphology (e.g.,
\citealt{lardo12}).  We have not considered these stars any further in the
analysis. 

\begin{table*}
 \centering
 \begin{minipage}{180mm}
  \caption{Program stars observed with AAOmega.} 
  \label{tab:aao} 
  \begin{tabular}{@{}llcccccrrrr@{}}
  \hline
	Name1\footnote{AXXX names are from \citet{arp55}, CRXXX names are from
\citet{cudworth87}, HXXX names are from \citet{harris75}, and NR XXX names are
from the \citet{grundahl99} photometry.}  & 
	Name2 & 
	RA.\ 2000 & 
	Dec.\ 2000 & 
	P(\%)\footnote{Probability of cluster membership from \citet{cudworth87}.} & 
	Flag\footnote{1 = stars which lie on the anomalous giant branch selected
from the $v$ versus $u-y$ CMD. All other stars lie on the canonical RGB.} &
	$V$ & 
	RV (\kms) & 
	$\sigma$RV (\kms) &
	$S(3839)$ & 
	$m_{\rm CH}$ \\
	(1) & 
	(2) &
	(3) &
	(4) & 
	(5) & 
	(6) & 
	(7) & 
	(8) & 
	(9) & 
	(10) & 
	(11) \\ 
  \hline
     \multicolumn{11}{c}{2010 09 30}   \\ 
  \hline
HI-240 & AIII-43       & 21 33 10.70 & $-$00 51 09.67 &     99 & \ldots &   14.25 &     1.6 & 2.0 & 1.183 &    0.082 \\
HI-451 & \ldots        & 21 33 39.11 & $-$00 49 30.18 & \ldots & \ldots &   15.86 &     2.1 & 4.6 & 0.574 &    0.150 \\
NR  38 & \ldots        & 21 33 28.91 & $-$00 50 00.94 & \ldots &      1 &   13.60 &     1.1 & 2.2 & 0.520 &    0.045 \\
NR  76 & HI-104,AII-30 & 21 33 17.91 & $-$00 48 19.82 &     99 & \ldots &   13.85 &  $-$1.5 & 4.9 & 0.291 &    0.049 \\
NR  81 & \ldots        & 21 33 27.08 & $-$00 48 19.41 & \ldots &      1 &   13.81 & $-$16.8 & 2.2 & 0.350 &    0.157 \\
NR 124 & \ldots        & 21 33 27.81 & $-$00 47 30.43 & \ldots & \ldots &   14.21 &     4.3 & 2.3 & 0.421 &    0.009 \\
NR 132 & \ldots        & 21 33 23.10 & $-$00 48 11.53 & \ldots &      1 &   14.20 &     0.8 & 2.5 & 0.245 &    0.023 \\
NR 184\footnote{\label{note1}NR 184 is a UV-bright star, NR 648 is a BHB star, NR 707 is a
RHB (or AGB) star and NR 82 is an AGB star.} & CR57          & 21 33 24.94 & $-$00 50 41.42 &     99 & \ldots &   15.24 &     0.0 & 7.8 & 0.002 & $-$0.113 \\
NR 216 & CR19          & 21 33 31.48 & $-$00 49 06.33 &     99 &      1 &   14.82 & $-$12.1 & 2.7 & 0.652 &    0.038 \\
NR 225 & HI-586,AI-58  & 21 33 29.27 & $-$00 45 55.49 &     99 & \ldots &   14.82 & $-$10.6 & 5.2 & 0.492 &    0.025 \\
NR 301 & AIV-37        & 21 33 32.80 & $-$00 50 27.06 &     99 &      1 &   14.93 &  $-$6.3 & 3.2 & 0.693 &    0.098 \\
NR 358 & HI-521,AI-79  & 21 33 34.05 & $-$00 47 32.10 &     99 &      1 &   15.06 & $-$16.0 & 4.2 & 1.459 &    0.043 \\
NR 378 & AI-50         & 21 33 30.32 & $-$00 47 24.54 &     99 &      1 &   15.22 &     3.2 & 5.2 & 0.363 &    0.086 \\
NR 386 & \ldots        & 21 33 26.18 & $-$00 49 21.35 & \ldots &      1 &   15.70 &  $-$3.4 & 3.7 & 0.057 & $-$0.061 \\
NR 388 & \ldots        & 21 33 27.16 & $-$00 50 25.43 & \ldots & \ldots &   15.23 &  $-$6.0 & 6.0 & 0.293 & $-$0.022 \\
NR 417 & CR76          & 21 33 23.48 & $-$00 48 46.57 &     99 & \ldots &   15.39 &  $-$2.9 & 3.3 & 0.212 &    0.017 \\
NR 721 & \ldots        & 21 33 24.32 & $-$00 49 41.46 & \ldots &      1 &   15.79 &  $-$8.8 & 1.8 & 0.309 &    0.002 \\
NR 811 & \ldots        & 21 33 22.85 & $-$00 50 34.00 & \ldots & \ldots &   15.74 &  $-$9.3 & 5.2 & 0.388 &    0.013 \\
NR 847 & \ldots        & 21 33 23.45 & $-$00 46 24.34 & \ldots & \ldots &   15.79 & $-$16.1 & 2.8 & 0.430 & $-$0.008 \\
NR 915 & \ldots        & 21 33 35.36 & $-$00 49 57.45 & \ldots & \ldots &   15.83 &  $-$5.2 & 6.9 & 0.319 &    0.037 \\
NR 1178 & \ldots        & 21 33 31.64 & $-$00 49 59.80 & \ldots &      1 &   16.06 &     3.0 & 5.1 & 0.343 &    0.059 \\
NR 1204 & AIII-26       & 21 33 20.08 & $-$00 50 13.76 &     99 &      1 &   15.82 &  $-$7.9 & 3.3 & 0.561 &    0.154 \\
  \hline
     \multicolumn{11}{c}{2011 11 01}   \\ 
  \hline
HI-240 & AIII-43       & 21 33 10.70 & $-$00 51 09.67 &     99 & \ldots &   14.25 &     2.9 & 2.5 & 1.069 &    0.085 \\
HI-451 & \ldots        & 21 33 39.11 & $-$00 49 30.18 & \ldots & \ldots &   15.86 &     4.3 & 2.7 & 0.609 &    0.199 \\
NR  47 & CR12          & 21 33 28.52 & $-$00 48 43.92 &     99 &      1 &   13.70 &     5.1 & 1.5 & 0.483 &    0.058 \\
NR  76 & HI-104,AII-30 & 21 33 17.91 & $-$00 48 19.82 &     99 & \ldots &   13.85 &  $-$1.5 & 1.9 & 0.283 &    0.045 \\
NR  82$^{\ref{note1}}$   & CR190         & 21 33 33.63 & $-$00 50 29.50 &     99 & \ldots &   13.91 &  $-$3.1 & 1.7 & 0.424 & $-$0.015 \\
NR  99 & AIII-86       & 21 33 23.59 & $-$00 50 41.07 &     99 & \ldots &   13.69 &  $-$1.3 & 2.1 & 0.250 &    0.018 \\
NR 124 & \ldots        & 21 33 27.81 & $-$00 47 30.43 & \ldots & \ldots &   14.21 &     4.4 & 1.9 & 0.382 &    0.017 \\
NR 132 & \ldots        & 21 33 23.10 & $-$00 48 11.53 & \ldots &      1 &   14.20 &     1.2 & 2.2 & 0.337 &    0.053 \\
NR 207 & \ldots        & 21 33 27.48 & $-$00 49 51.35 & \ldots &      1 &   14.89 &  $-$1.1 & 2.0 & 0.219 &    0.004 \\
NR 225 & HI-586,AI-58  & 21 33 29.27 & $-$00 45 55.49 &     99 & \ldots &   14.82 & $-$10.8 & 2.0 & 0.430 &    0.049 \\
NR 254 & \ldots        & 21 33 29.37 & $-$00 49 42.84 & \ldots &      1 &   15.05 &     3.4 & 2.0 & 0.255 &    0.027 \\
NR 299 & AI-22         & 21 33 35.32 & $-$00 49 22.13 &     99 & \ldots &   14.94 &     0.3 & 1.9 & 0.307 &    0.120 \\
NR 358 & HI-521,AI-79  & 21 33 34.05 & $-$00 47 32.10 &     99 &      1 &   15.06 & $-$16.7 & 1.2 & 1.328 &    0.114 \\
NR 375 & \ldots        & 21 33 30.62 & $-$00 50 08.33 & \ldots & \ldots &   15.20 &     2.1 & 1.7 & 0.322 &    0.022 \\
NR 378 & AI-50         & 21 33 30.32 & $-$00 47 24.54 &     99 &      1 &   15.22 &  $-$2.3 & 2.0 & 0.432 &    0.081 \\
NR 403 & CR58          & 21 33 25.64 & $-$00 50 43.12 &     99 & \ldots &   15.20 & $-$15.3 & 1.8 & 0.331 &    0.015 \\
NR 648$^{\ref{note1}}$ & \ldots        & 21 33 26.32 & $-$00 49 10.58 & \ldots & \ldots &   15.99 &  $-$1.6 & 1.9 & 0.194 & $-$0.049 \\
NR 707$^{\ref{note1}}$ & \ldots        & 21 33 23.02 & $-$00 48 56.69 & \ldots & \ldots &   15.81 &     0.3 & 1.9 & 0.095 & $-$0.053 \\
NR 801 & \ldots        & 21 33 27.46 & $-$00 46 53.10 & \ldots & \ldots &   15.79 &  $-$2.3 & 2.2 & 0.305 & $-$0.004 \\
NR 847 & \ldots        & 21 33 23.45 & $-$00 46 24.34 & \ldots & \ldots &   15.79 &  $-$3.9 & 2.0 & 0.377 &    0.034 \\
NR 915 & \ldots        & 21 33 35.36 & $-$00 49 57.45 & \ldots & \ldots &   15.83 & $-$11.8 & 2.3 & 0.332 &    0.047 \\
NR 947 & AIII-10       & 21 33 18.75 & $-$00 49 44.09 &     99 & \ldots &   15.79 &  $-$6.2 & 1.9 & 0.190 &    0.059 \\
NR 1204 & AIII-26       & 21 33 20.08 & $-$00 50 13.76 &     99 &      1 &   15.82 & $-$10.6 & 1.7 & 0.449 &    0.123 \\
\hline
\end{tabular}
\end{minipage}
\end{table*}

\subsection{High-resolution spectroscopic observations} 

Three stars (NR 76, NR 81 and NR 132) were observed in service mode using the
High Dispersion Spectrograph (HDS, \citealt{noguchi02}) on the Subaru Telescope
on 2011 August 3. Six additional stars (NR 47, NR 99, NR 124, NR 207, NR 254
and NR 378) were observed using HDS in classical mode on 2013 July 17.  All
nine stars were also observed with the AAOmega instrument. For both sets of
observations, we used the StdYb setting and the 0\farcs8 slit which resulted in
a wavelength coverage from $\sim$4100\,\AA\ to $\sim$6800\,\AA\ at a spectral
resolution of $R = 45000$.  A telluric standard was also observed.  The
spectra were reduced using {\sc iraf}\footnote{{\sc iraf} is distributed by the
National Optical Astronomy Observatories, which are operated by the Association
of Universities for Research in Astronomy, Inc., under cooperative agreement
with the National Science Foundation.} adopting a similar approach as in
\citet{yong06}. 

Five stars (NR~37, NR~38, NR~58, NR~60 and NR~77) were also observed using the
Magellan Inamori Kyocera Echelle (MIKE) spectrograph \citep{bernstein03} at the
Magellan Telescope on 2012 August 26.  NR 60 is a likely AGB star based on CMD
location.  Full wavelength coverage was obtained ($\sim$3400\,\AA\ to
$\sim$9000\,\AA), and we used the 0\farcs7 slit which provided a spectral
resolution of $R = 40000$ in the blue arm and $R = 35000$ in the red arm, as
measured from the ThAr lines. The spectra were reduced using the {\sc CarPy} 
pipeline\footnote{http://code.obs.carnegiescience.edu/mike} and independently
in {\sc iraf} using the {\sc mtools} 
package\footnote{http://www.lco.cl/telescopes-information/magellan/instruments/mike/iraf-tools/iraf-mtools-package}.
One star, NR 77, had a faint, nearby companion.  This object was reduced in two
different ways using {\sc iraf}. In the first approach, we adopted a
conservative aperture placement to try to avoid flux from the faint companion.
In the second approach, the flux from both stars was extracted. In the
subsequent section, we analyze both sets of spectra independently in order to
quantify the contamination from the nearby companion.  The program stars are
listed in Table \ref{tab:param}. We note that of these 14 objects observed at
high spectral resolution, six belong to the canonical RGB and eight are
anomalous RGB stars. 

\begin{table*}
 \centering
 \begin{minipage}{180mm}
  \caption{Program stars and stellar parameters for objects observed with
Magellan or Subaru.} 
  \label{tab:param} 
  \begin{tabular}{@{}llccccccrcccccc@{}}
  \hline
         Name1\footnote{AXXX names are from \citet{arp55}, CRXXX names are from
\citet{cudworth87}, HXXX names are from \citet{harris75}, and NR XXX names are
from the \citet{grundahl99} photometry.}	& 
         Name2	& 
         R.A. & 
         Dec. & 
         P\footnote{Probability of cluster membership from
\citet{cudworth87}.}        &
         Flag\footnote{1 = stars which lie on the anomalous giant branch selected
from the $v$ versus $u-y$ CMD. All other stars lie on the canonical RGB.} &
	 Run\footnote{M12 = Magellan Telescope 2012 08 26, S11 =
Subaru Telescope 2011 08 03, S13 = Subaru Telescope 2013 07 17.} & 
         $V$ & 
         RV & 
         $\sigma$RV &
         \teff & 
         \logg &
         \vt & 
         [m/H]\footnote{[m/H] refers to the metallicity used to generate
the model atmosphere.} & 
         [Fe/H] \\
         & 
         & 
         & 
         & (\%) 
         & 
         & 
         & 
         & 
\multicolumn{2}{c}{(\kms)} &  
         (K) &
         (cgs) & 
         (\kms) & 
         (dex) & 
         (dex) \\ 
         (1) & 
         (2) &
         (3) &
         (4) & 
         (5) & 
         (6) &
         (7) & 
         (8) & 
         (9) &
         (10) &
         (11) &
         (12) & 
         (13) & 
         (14) &
         (15) \\ 
  \hline
\multicolumn{14}{c}{Canonical RGB ($r$-only) stars (black circles or lines in the figures)} \\
  \hline 
NR  37  & CR78          & 21 33 25.44 & $-$00 48 53.73 &     99 & \ldots & M12 & 13.56 & $-$15.3 & 1.0 & 4250 & 0.70 & 1.77 & $-$1.6 & $-$1.66 \\ 
NR  58  & CR30          & 21 33 32.17 & $-$00 50 01.17 &     99 & \ldots & M12 & 13.55 &    11.8 & 1.0 & 4225 & 0.70 & 1.89 & $-$1.6 & $-$1.64 \\ 
NR  60\footnote{NR 60 is a likely AGB star.}  & CR28          & 21 33 32.57 & $-$00 49 45.72 &     99 & \ldots & M12 & 13.55 &  $-$7.1 & 1.0 & 4325 & 0.30 & 2.19 & $-$1.7 & $-$1.75 \\ 
NR  76  & HI-104        & 21 33 17.91 & $-$00 48 19.82 &     99 & \ldots & S11 & 13.85 &  $-$1.3 & 0.6 & 4375 & 0.90 & 1.73 & $-$1.7 & $-$1.69 \\ 
NR  99  & AIII-86       & 21 33 23.59 & $-$00 50 41.07 &     99 & \ldots & S13 & 13.69 &  $-$1.5 & 0.6 & 4275 & 0.70 & 1.78 & $-$1.6 & $-$1.66 \\ 
NR 124  & \ldots        & 21 33 27.81 & $-$00 47 30.43 & \ldots & \ldots & S13 & 14.21 &     3.4 & 0.7 & 4425 & 0.85 & 1.81 & $-$1.6 & $-$1.64 \\ 
  \hline
\multicolumn{14}{c}{Anomalous RGB ($r+s$) stars with [Fe/H] $\approx$ $-$1.5 (red triangles or lines in the figures)} \\
  \hline 
NR  38  & \ldots        & 21 33 28.91 & $-$00 50 00.94 & \ldots &      1 & M12 & 13.60 &     3.7 & 1.3 & 4175 & 0.60 & 2.12 & $-$1.6 & $-$1.61 \\ 
NR  47  & CR12          & 21 33 28.52 & $-$00 48 43.92 &     99 &      1 & S13 & 13.70 &     3.3 & 0.5 & 4050 & 0.65 & 1.77 & $-$1.4 & $-$1.42 \\ 
NR  77  & \ldots        & 21 33 24.45 & $-$00 48 36.29 & \ldots &      1 & M12 & 13.92 &     6.6 & 1.0 & 4350 & 1.00 & 2.25 & $-$1.5 & $-$1.46 \\ 
NR  81  & \ldots        & 21 33 27.08 & $-$00 48 19.41 & \ldots &      1 & S11 & 13.81 & $-$22.0 & 0.5 & 4275 & 1.00 & 1.85 & $-$1.6 & $-$1.55 \\ 
  \hline
\multicolumn{14}{c}{Anomalous RGB (metal-rich) stars with [Fe/H] $\approx$ $-$1.0 (aqua star symbols or lines in the figures)} \\
  \hline 
NR 132  & \ldots        & 21 33 23.10 & $-$00 48 11.53 & \ldots &      1 & S11 & 14.20 &     0.7 & 0.5 & 4325 & 1.30 & 1.88 & $-$1.0 & $-$0.97 \\ 
NR 207  & \ldots        & 21 33 27.48 & $-$00 49 51.35 & \ldots &      1 & S13 & 14.89 &  $-$2.1 & 0.4 & 4425 & 1.30 & 1.40 & $-$1.1 & $-$1.08 \\ 
NR 254  & \ldots        & 21 33 29.37 & $-$00 49 42.84 & \ldots &      1 & S13 & 15.05 &     3.2 & 0.5 & 4525 & 1.60 & 1.61 & $-$1.0 & $-$0.97 \\ 
NR 378  & AI-50         & 21 33 30.32 & $-$00 47 24.54 &     99 &      1 & S13 & 15.22 &  $-$2.9 & 0.5 & 4750 & 1.50 & 1.68 & $-$1.1 & $-$1.08 \\ 
  \hline
\hline
\end{tabular}
\end{minipage}
\end{table*}

\begin{figure}
\centering
      \includegraphics[width=.99\hsize]{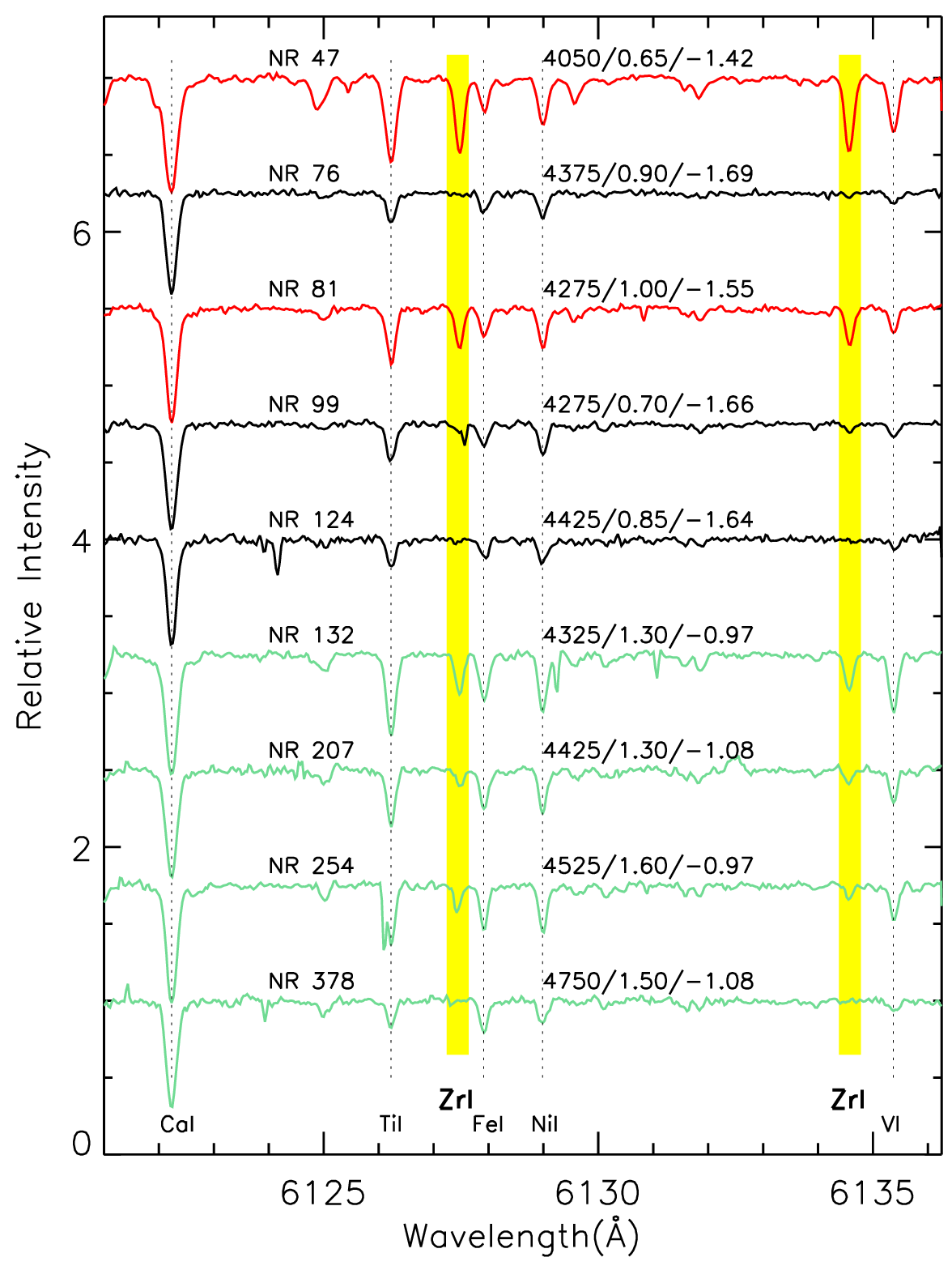}
      \caption{A portion of the Subaru HDS spectra for nine program stars. The
yellow region highlights \zri\ lines used in the analysis. There are clear
star-to-star differences in the \zri\ line strengths, and also for \vi\ and
\tii\ lines. Black lines represent stars that lie on the canonical RGB. Red
lines are stars on the anomalous RGB. The aqua lines are the unusually
metal-rich objects on the anomalous RGB. (The colours are consistent with those
used in Figure \ref{fig:cmd100}.) The positions of other atomic lines and the
stellar parameters (\teff/\logg/[Fe/H]) are included. 
      \label{fig:spec1s} }
\end{figure}

\begin{figure}
\centering
      \includegraphics[width=.99\hsize]{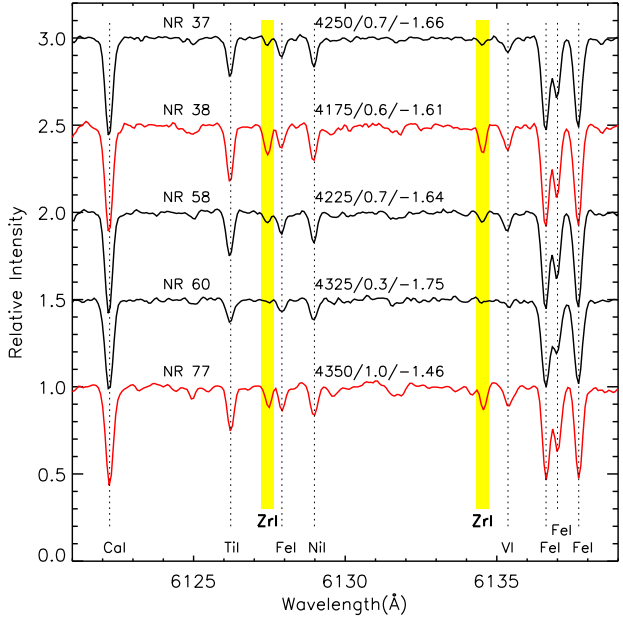}
      \caption{A portion of the Magellan MIKE spectra for five program stars. As in
Figure \ref{fig:spec1s}, Zr lines are highlighted and there are significant
star-to-star differences in line strengths. The black spectra denote that those
stars lie on the canonical RGB while red spectra represent stars that lie on
the anomalous RGB. The positions of other atomic lines and the stellar
parameters (\teff/\logg/[Fe/H]) are included. 
      \label{fig:spec1m} }
\end{figure}

\begin{figure}
\centering
      \includegraphics[width=.99\hsize]{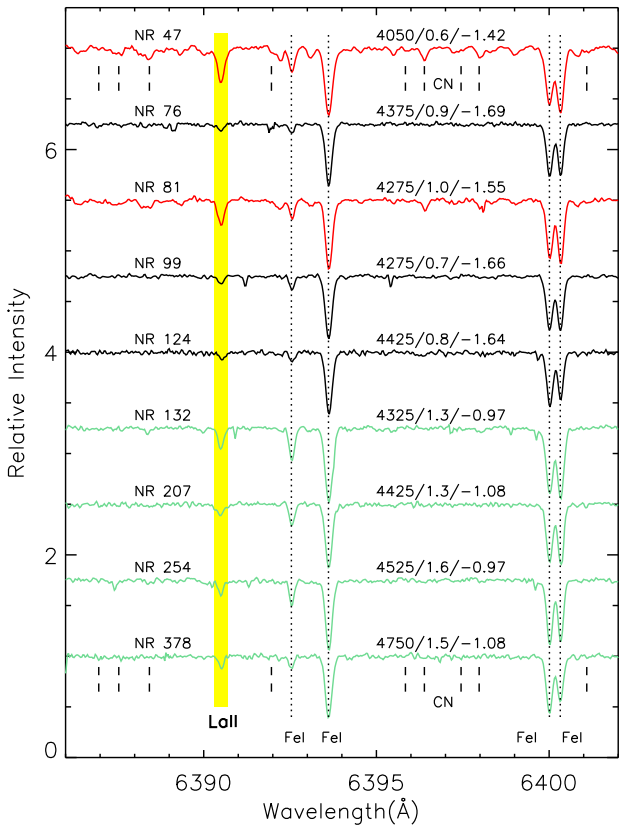}
      \caption{Same as Figure \ref{fig:spec1s} but for a region containing a La line
used in the analysis. There are significant star-to-star differences in the
line strength of La. The positions of CN lines are marked. 
      \label{fig:spec2s} }
\end{figure}

\begin{figure}
\centering
      \includegraphics[width=.99\hsize]{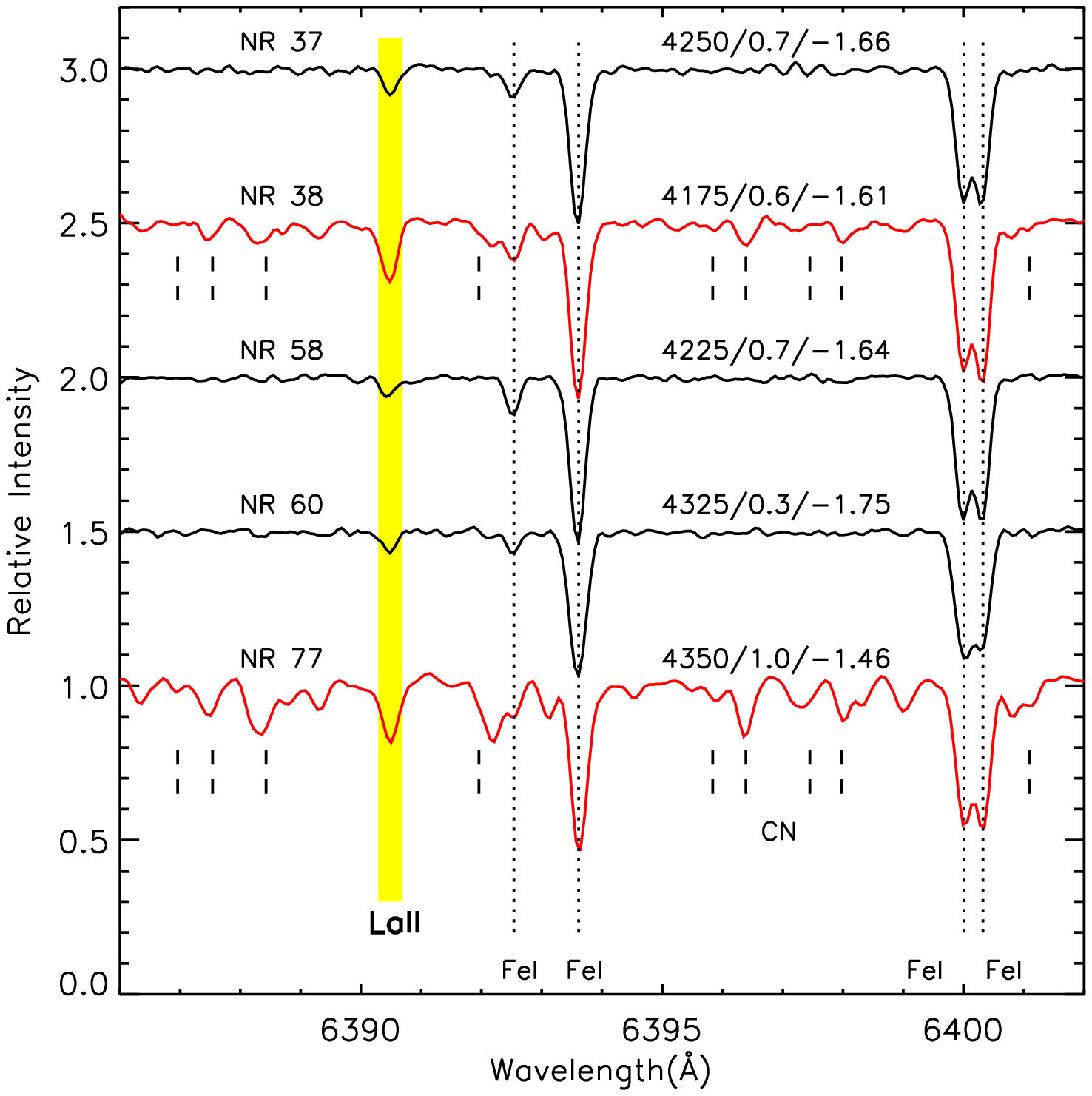}
      \caption{Same as Figure \ref{fig:spec2s} but for the Magellan MIKE
spectra. 
      \label{fig:spec2m} }
\end{figure}

In Figures \ref{fig:spec1s} to \ref{fig:spec2m}, we plot regions of the high
dispersion spectra for the program stars from both telescope+instrument
combinations. These figures demonstrate that there are star-to-star variations
in the strengths of Zr and La lines which could be caused by differences in
stellar parameters and/or chemical abundance ratios. In Section 3.2, we shall
seek to quantify the stellar parameters and chemical abundances. 

\section{ANALYSIS}

\subsection{Radial velocities and line indices from medium-resolution spectra}

Radial velocities were measured from the medium-resolution spectra by
cross-correlating each program star against HI-240. The radial velocity for
HI-240 was determined by measuring the wavelengths of a small set of lines
(sodium doublet 5889.951\,\AA\ and 5895.924\,\AA\ and the calcium triplet
8498.03\,\AA, 8542.09\,\AA\ and 8662.14\,\AA). Given the superior spectral
resolution in the red arm, we adopted those values as the radial velocities and
corrected for the heliocentric motion. We measured the \scn\ and \mch\ indices
in the AAOmega spectra using the definitions given in \citet{smith90}. Calcium
triplet line strengths were measured via Gaussian line profile fits to the
observed data for the two stronger \caii\ triplet lines at 8542\,\AA\ and
8662\,\AA\ using the technique first described in \citet{AD91}.  The
heliocentric radial velocities and \scn\ and \mch\ indices are presented in
Table \ref{tab:aao}. 

An assessment of the internal errors associated with these measurement can be
obtained by consideration of the 11 objects observed on both runs. For the
radial velocities, 10 of the 11 objects showed no evidence ($\le$1-$\sigma$,
i.e., less than one standard deviation) for radial velocity variation between
the two observing runs. One star, NR 847, exhibited evidence for radial
velocity variability; $-$16.1 \kms\ $\pm$ 2.8 \kms\ (2010) versus $-$3.9 \kms\
$\pm$ 2.0 \kms\ (2011). Excluding NR 847, the average difference in radial
velocity for stars observed on both runs is 1.2 \kms\ $\pm$ 0.9 \kms\ ($\sigma$
= 2.9 \kms). For the \scn\ and \mch\ indices, we find mean differences for
stars observed on both runs of 0.028 $\pm$ 0.023 ($\sigma$ = 0.075) and 0.018
$\pm$ 0.009 ($\sigma$ = 0.029), respectively.  Since for the \caii\ triplet
spectra only single observations are available, we adopt the uncertainty in the
pseudo-equivalent widths which results from the uncertainties in the Gaussian
fit parameters for the observed line profiles. 

An assessment of the systematic errors can be obtained by comparison of our
measurements with literature values. For the radial velocities, five of our
program stars were also observed by \citet{lardo12,lardo13}, noting that on
average their measurement errors ($\langle\sigma$RV$\rangle$ = 16.5 \kms) are
larger than ours ($\langle\sigma$RV$\rangle$ = 3.0 \kms). For three of these
five stars our radial velocity measurements are in agreement. The two stars
with poor agreement are NR 132, $-$16.4 \kms\ $\pm$ 7 \kms\ \citep{lardo13}
versus 0.9 \kms\ $\pm$ 3.4 \kms\ (this study), and NR 378, $-$60.3 \kms\ $\pm$
5.8 \kms\ \citep{lardo13} versus 0.5 \kms\ $\pm$ 5.6 \kms\ (this study). 
These stars may be spectroscopic binaries. 

The \scn\ and \mch\ values are in good agreement with those of \citet{smith90}
for the two stars in common. For HI-240, our mean values are \scn\ = 1.126 and
\mch\ = 0.084 and the \citet{smith90} values are 1.111 and 0.067, respectively.
For HI-451, our mean values are \scn\ = 0.592 and \mch\ = 0.175 and the
\citet{smith90} values are 0.571 and 0.165, respectively. We note that the
differences for \scn\ and \mch\ between this study and \citet{smith90} are
comparable to mean differences for the 11 stars observed on both AAOmega runs. 

\subsection{Stellar parameters, chemical abundances and radial velocities from
high-resolution spectra} 

Equivalent widths (EW) were measured from the high-resolution spectra using
routines in {\sc iraf} and {\sc daospec} \citep{stetson08}. When using
{\sc iraf} to measure an EW, every line in every star was visually inspected.
In a given star, lines regarded to be blended or poorly fit were excluded, and  
weak (EW $<$ 5 m\AA) and strong (EW $>$ 130 m\AA) lines were also removed from
the analysis. When using {\sc daospec} to measure EWs, the continuum was the
same as in the {\sc iraf} analysis, i.e., {\sc daospec} did not re-adjust the
continuum level.  Additionally, the set of lines measured using {\sc daospec}
was identical to those already measured, and visually inspected, using {\sc
iraf}. 
For the Subaru and
Magellan spectra, there was good agreement  ($\sigma$ = 1.5 m\AA) between
the two sets of EW measurements, and we adopted the {\sc daospec} values. For
the Magellan spectra, EW measurements could be compared between the {\sc CarPy}
reduction and the {\sc iraf} reduction. Again, there was excellent agreement
between the two sets of measurements ($\sigma$ = 1.4 m\AA). The EW measurements
and line list are presented in Table \ref{tab:ew}. 

\begin{table*}
 \centering
 \begin{minipage}{180mm}
  \caption{Line list for the program stars\label{tab:ew}}
  \begin{tabular}{@{}cccrrrrrrrrrrrrrrrc@{}}
  \hline
Wavelength & 
Species\footnote{The digits to the left of the decimal point are the atomic
number. The digit to the right of the decimal point is the ionization state
(``0'' = neutral, ``1'' = singly ionised).} & 
L.E.P & 
$\log gf$ & 
NR 37 &
NR 38 & 
NR 47 & 
NR 58 & 
NR 60 & 
NR 76 & 
NR 77 & 
NR 81 & 
Source\footnote{A = $\log gf$ values taken from \citet{yong05} where the references include 
\citet{denhartog03}, \citet{ivans01}, \citet{kurucz95}, 
\citet{prochaska00}, \citet{ramirez02}; 
B = \citet{gratton03};
C = Oxford group including 
\citet{blackwell79feb,blackwell79fea,blackwell80fea,blackwell86fea,blackwell95fea}; 
D = \citet{biemont91}; 
E1 = \citet{fuhr09}, using line component patterns for 
     hfs/IS from \citet{kurucz95}; 
E2 = \citet{roederer12c};
E3 = \citet{fuhr09};
E4 = \citet{biemont11};
E5 = \citet{biemont81};
E6 = \citet{ljung06};
E7 = \citet{whaling88};
E8 = \citet{fuhr09}, using hfs/IS from \citet{mcwilliam98};
E9 = \citet{lawler01a}, using hfs from \citet{ivans06};
E10 = \citet{lawler09};
E11 = \citet{li07}, using hfs from \citet{sneden09};
E12 = \citet{ivarsson01}, using hfs from \citet{sneden09};
E13 = \citet{denhartog03}, using hfs/IS from \citet{roederer08} when available;
E14 = \citet{lawler06}, using hfs/IS from \citet{roederer08} when available;
E15 = \citet{lawler01b}, using hfs/IS from \citet{ivans06};
E16 = \citet{roederer12b};
E17 = \citet{denhartog06};
E18 = \citet{lawler01c}, using hfs from \citet{lawler01d,lawler09};
E19 = \citet{wickliffe00};
E20 = \citet{lawler08};
E21 = \citet{wickliffe97};
E22 = \citet{sneden09} for \loggf\ and hfs/IS;
E23 = \citet{lawler07};
E24 = \citet{biemont00}, using hfs/IS from \citet{roederer12b}.
\\
This table is published in its entirety in the electronic edition of the paper.
A portion is shown here for guidance regarding its form and content.} \\ 
\AA & 
 & 
eV & 
 & 
m\AA & 
m\AA & 
m\AA & 
m\AA & 
m\AA & 
m\AA & 
m\AA & 
m\AA & 
 \\ 
(1) & 
(2) &
(3) &
(4) & 
(5) & 
(6) & 
(7) & 
(8) & 
(9) & 
(10) &
(11) & 
(12) & 
(13) \\ 
\hline
 6300.31 &    8.0 &   0.00 & $-$9.75 &     32.3 &     53.5 &   \ldots &     43.7 &     23.3 &   \ldots &   \ldots &     49.3 &           B \\
 6363.78 &    8.0 &   0.02 &$-$10.25 &   \ldots &     26.3 &   \ldots &     16.1 &     12.0 &   \ldots &     14.2 &     22.3 &           A \\
 4751.82 &   11.0 &   2.10 & $-$2.11 &   \ldots &   \ldots &   \ldots &   \ldots &   \ldots &   \ldots &     11.2 &   \ldots &           B \\
 4982.83 &   11.0 &   2.10 & $-$0.91 &   \ldots &   \ldots &   \ldots &   \ldots &   \ldots &     20.8 &     48.2 &     39.0 &           A \\
 5682.65 &   11.0 &   2.10 & $-$0.67 &     51.6 &     49.7 &    112.7 &     37.8 &     45.5 &     24.9 &   \ldots &     57.6 &           B \\ 
\hline
\end{tabular}
\end{minipage}
\end{table*}

To determine the stellar parameters, we adopted a traditional spectroscopic
approach. We used one dimensional local thermodynamic equilibrium (LTE) model
atmospheres  with [$\alpha$/Fe] = +0.4 from the \citet{castelli03} grid.  To
produce particular models we used the interpolation software tested in
\citet{allende04}. Chemical abundances were computed using the LTE stellar line
analysis program {\sc moog} \citep{moog,sobeck11}. The effective temperature, \teff,
was adjusted until there was no trend between the abundance from \fei\ lines
and the lower excitation potential (L.E.P). The surface gravity, \logg, was
adjusted until the abundance from \fei\ lines was the same as from \feii\
lines.  The microturbulent velocity, \vt, was established when there was no
trend between the abundance from \fei\ and the reduced equivalent width, EW$_r$
= $\log ($EW$/\lambda)$. Finally, we required that the derived metallicity was
within 0.1 dex of the value adopted in the model atmosphere.  The final stellar
parameters (see Table \ref{tab:param}) were obtained when these four conditions
were simultaneously satisfied. We note that NR 60, whose CMD location is
consistent with being an AGB star, has a surface gravity appropriate for that
evolutionary phase. 

Uncertaintes in the stellar parameters were obtained in the following
manner. For \teff\ and \vt\, we measured the uncertainty in the slope between
the abundance from \fei\ lines and L.E.P.\ and EW$_r$, respectively. We then
adjusted \teff\ or \vt\ until the slope matched the relevant uncertainty. For
\logg, we added the standard error of the mean for \fei\ and \feii\, in
quadrature, then adjusted \logg\ until the difference in abundances from \fei\
and \feii\ was equal to this value. Adopting this approach, we estimate that
the internal uncertainties in \teff, \logg\ and \vt\ are 50 K, 0.2 dex and
0.2 \kms, respectively, and these are slightly conservative estimates. 

For \teff\ and \logg, we can compare the spectroscopic values to photometric
values. For \teff, we used the infrared flux method (IRFM)
metallicity-dependent colour-temperature relations of \citet{ramirez05} for
giant stars. We assumed a reddening $E(B-V)$ = 0.06 as in the \citet{harris96}
catalogue\footnote{Here and throughout the paper, we use the values found in
the 2010 version of the catalogue (available online) rather than the values in
the original \citet{harris96} paper.}. The values are the weighted mean from
the $b-y$, $V-J$, $V-H$ and $V-K$ colours ($JHK$ photometry from 2MASS,
\citealt{2mass}).  The surface gravity was determined assuming the photometric
\teff, a distance modulus $(m-M)_V$ = 15.5 \citep{harris96}, bolometric
corrections from \citet{alonso99} and a mass of 0.8 \msun. The mean differences
(photometric $-$ spectroscopic) in \teff\ and \logg\ are $-$13 $\pm$ 26 K
($\sigma$ = 78 K) and +0.08 $\pm$ 0.07 dex ($\sigma$ = 0.20 dex), respectively.
These differences are within the uncertainties estimated above. 

\begin{figure}
\centering
      \includegraphics[width=.99\hsize]{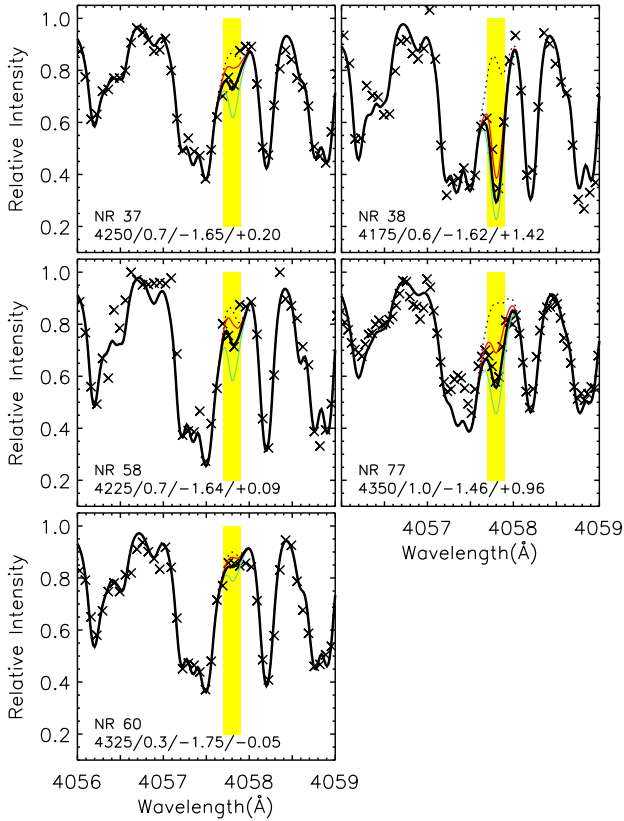}
      \caption{A portion of the Magellan MIKE spectra near the 4057\,\AA\ \pbi\  
line, highlighted in yellow. The black thick line represents the best-fitting
synthetic spectra. The red and aqua lines show synthetic spectra with
unsatisfactory ratios of [Pb/Fe], namely $\Delta$Pb $\pm$ 0.5 dex from the line
of best fit. The dotted black line is a synthesis containing no Pb. The values
written in the bottom of each panel are \teff/\logg/[Fe/H]/[Pb/Fe]. 
      \label{fig:synthpb} }
\end{figure}

For Ni and lighter elements, chemical abundances were computed using the
measured EWs, final model atmospheres and {\sc moog}. For Cu, Zn and the
neutron-capture elements, abundances were determined via spectrum synthesis
(e.g., see Figure \ref{fig:synthpb} for the Pb analysis). Lines affected by
hyperfine splitting (hfs) and/or isotope shifts (IS) were treated appropriately
using the hfs data from \citet{kurucz95} or other sources as noted in Table
\ref{tab:ew}. We adopted the \citet{asplund09} solar abundances.  The chemical
abundances are presented in Table \ref{tab:abun}. 

\begin{table}
 \centering
 \begin{minipage}{90mm}
  \caption{Chemical abundances for the program stars.}
  \label{tab:abun} 
  \begin{tabular}{@{}lcrcrc@{}}
  \hline
	Name & 
	A(X) & 
	N$_{\rm lines}$ & 
	s.e.$_{\log\epsilon}$ & 
	[X/Fe] & 
	Total Error \\
\hline 
\hline
     \multicolumn{6}{c}{\oi}   \\ 
\hline
NR 37   &    7.48 & 1      &  \ldots &    0.44 &    0.23 \\
NR 38   &    7.71 & 2      &    0.06 &    0.64 &    0.19 \\
NR 47   &  \ldots & \ldots &  \ldots &  \ldots &  \ldots \\
NR 58   &    7.62 & 2      &    0.00 &    0.57 &    0.19 \\
NR 60   &    7.20 & 2      &    0.10 &    0.26 &    0.18 \\
NR 76   &  \ldots & \ldots &  \ldots &  \ldots &  \ldots \\
NR 77   &  \ldots & \ldots &  \ldots &  \ldots &  \ldots \\
NR 81   &    7.83 & 2      &    0.03 &    0.69 &    0.18 \\
NR 99   &  \ldots & \ldots &  \ldots &  \ldots &  \ldots \\
NR 124  &  \ldots & \ldots &  \ldots &  \ldots &  \ldots \\
NR 132  &  \ldots & \ldots &  \ldots &  \ldots &  \ldots \\
NR 207  &  \ldots & \ldots &  \ldots &  \ldots &  \ldots \\
NR 254  &  \ldots & \ldots &  \ldots &  \ldots &  \ldots \\
NR 378  &    7.97 & 1      &  \ldots &    0.36 &    0.24 \\
\hline
     \multicolumn{6}{c}{\nai}   \\ 
\hline
NR 37   &    4.76 & 3      &    0.03 &    0.18 &    0.13 \\
NR 38   &    4.73 & 3      &    0.06 &    0.11 &    0.13 \\
NR 47   &    5.44 & 3      &    0.01 &    0.63 &    0.14 \\
NR 58   &    4.44 & 3      &    0.03 & $-$0.16 &    0.13 \\
NR 60   &    4.84 & 3      &    0.06 &    0.35 &    0.13 \\
NR 76   &    4.43 & 3      &    0.06 & $-$0.12 &    0.13 \\
NR 77   &    5.29 & 4      &    0.10 &    0.52 &    0.12 \\
NR 81   &    4.84 & 5      &    0.02 &    0.15 &    0.11 \\
NR 99   &    4.43 & 2      &    0.06 & $-$0.14 &    0.15 \\
NR 124  &    4.86 & 3      &    0.01 &    0.26 &    0.13 \\
NR 132  &    5.14 & 5      &    0.04 & $-$0.13 &    0.10 \\
NR 207  &    5.01 & 4      &    0.04 & $-$0.14 &    0.11 \\
NR 254  &    5.08 & 4      &    0.03 & $-$0.18 &    0.11 \\
NR 378  &    4.93 & 2      &    0.03 & $-$0.22 &    0.15 \\
\hline
     \multicolumn{6}{c}{\mgi}   \\ 
\hline
NR 37   &    6.27 & 4      &    0.03 &    0.33 &    0.11 \\
NR 38   &    6.55 & 4      &    0.06 &    0.56 &    0.11 \\
NR 47   &  \ldots & \ldots &  \ldots &  \ldots &  \ldots \\
NR 58   &    6.42 & 5      &    0.04 &    0.47 &    0.10 \\
NR 60   &    6.27 & 3      &    0.01 &    0.42 &    0.13 \\
NR 76   &    6.23 & 3      &    0.02 &    0.32 &    0.12 \\
NR 77   &    6.59 & 4      &    0.05 &    0.46 &    0.11 \\
NR 81   &    6.40 & 3      &    0.05 &    0.35 &    0.12 \\
NR 99   &    6.38 & 3      &    0.10 &    0.45 &    0.13 \\
NR 124  &    6.25 & 2      &    0.03 &    0.29 &    0.15 \\
NR 132  &    6.86 & 4      &    0.04 &    0.22 &    0.11 \\
NR 207  &    6.79 & 1      &  \ldots &    0.28 &    0.22 \\
NR 254  &    6.89 & 1      &  \ldots &    0.26 &    0.22 \\
NR 378  &    6.75 & 2      &    0.02 &    0.23 &    0.15 \\
\hline
\end{tabular}
\end{minipage}
This table is published in its entirety in the electronic edition of the paper.
A portion is shown here for guidance regarding its form and content. 
\end{table}

To determine the errors in chemical abundances, we repeated the analysis
varying the stellar parameters, one at a time, by the relevant uncertainties
noted above.  Additionally, we also changed the metallicity in the model,
[m/H], by 0.2 dex.  We added these four error terms in quadrature to obtain the
systematic uncertainty. We replaced the random error (s.e.$_{\log \epsilon}$)
by max(s.e.$_{\log \epsilon}$, 0.20/$\sqrt{N_{\rm lines}}$) where the second
term is what would be expected for a set of $N_{\rm lines}$ with a dispersion
of 0.20 dex (a conservative value based on the abundance dispersion exhibited
by \fei\ lines). To obtain the total error (presented in Table \ref{tab:abun}),
we added the systematic and random errors in quadrature. 

As noted in Section 2.2, star NR 77 had a faint companion and we extracted the
spectrum for this star in two different ways. In the first approach, we placed
the apertures for each order in such a way as to avoid flux from the faint
companion. In the second approach, we extracted the flux from both stars. The
stellar parameters and chemical abundances were essentially identical between
the two approaches. We present the values from the first approach and are
confident that the results for this star are not affected by contamination from
the faint companion. 

For the 14 stars observed at high spectral resolution, radial velocities
were obtained from the observed wavelengths of the lines used in the
equivalent-width analysis. Heliocentric corrections were applied and the radial
velocities are presented in Table \ref{tab:param}. For the ten stars observed
at high- and medium-resolution, we find an average radial velocity difference
(high-resolution $-$ medium-resolution) of $-$1.0 $\pm$ 0.7 \kms\ ($\sigma$ =
2.1 \kms). This agreement gives us additional confidence in our heliocentric
radial velocity measurements. 

\section{RESULTS} 

\subsection{Cluster membership} 

Cluster membership for any given star can be established through a combination
of the following criteria: ($i$) evolutionary status, ($ii$) location in CMD,
($iii$) radial velocity, ($iv$) distance from cluster center and ($v$) proper
motions. Regarding point ($i$), all stars selected from \strom\ CMDs have
colours and magnitudes consistent with being giant stars at the distance of M2.
In particular, all 14 stars observed with the Subaru Telescope or Magellan
Telescope are red giants with magnitudes consistent with the distance modulus
of M2. Concerning point ($ii$), all stars occupy plausible locations in all
CMDs (although we shall revisit this aspect in Section 4.4 taking into account
the derived metallicities).  Regarding point ($iii$), the heliocentric radial
velocity of M2 is $-$5.3 \kms\ $\pm$ 2 \kms\ and the central velocity
dispersion is 8.2 \kms\ $\pm$ 0.6 \kms\ \citep{harris96}.  While all stars have
a radial velocity consistent with cluster membership, the small value means
that radial velocity alone cannot confirm cluster membership.  Concerning point
($iv$), we note that all stars lie within the tidal radius (21\farcm45,
\citealt{harris96}). For point ($v$), proper motions, and membership
probabilities based on those measurements, were published by
\citet{cudworth87}. For the 16 stars with proper motion measurements, we note
that all are high probability cluster members, $P$ = 99\%. 

Whether or not the four anomalous RGB stars with [Fe/H] $\approx$ $-$1.0 are
cluster members obviously affects our conclusions. We remain open to both
possibilities, i.e., that these four stars may, or may not, be members.  That
said, in an upcoming study by Milone et al.\ (in preparation), recent \hst\
photometry reveals that the four metal-rich stars appear to lie on a narrow
well-defined RGB sequence that can be traced to the subgiant branch and main
sequence regions supporting the case for cluster membership. 

\subsection{Radial velocity and velocity dispersion} 

To determine the radial velocity and velocity dispersion for M2, we took the
following approach. We exclude NR~847 as this star exhibits radial velocity
variation. For stars with multiple radial velocity measurements, we adopt the
weighted mean for a given star. Assuming all stars are cluster members, we find
that the heliocentric radial velocity for M2 is $-$3.9 $\pm$ 1.1 \kms\ 
($\sigma$ = 7.0 \kms)\footnote{This value is the observed dispersion and is not
corrected for the contribution from velocity errors.}. These values are in
good agreement with those listed in the \citet{harris96} catalogue. 

\subsection{CN and CH indices} 

\begin{figure}
\centering
      \includegraphics[width=.99\hsize]{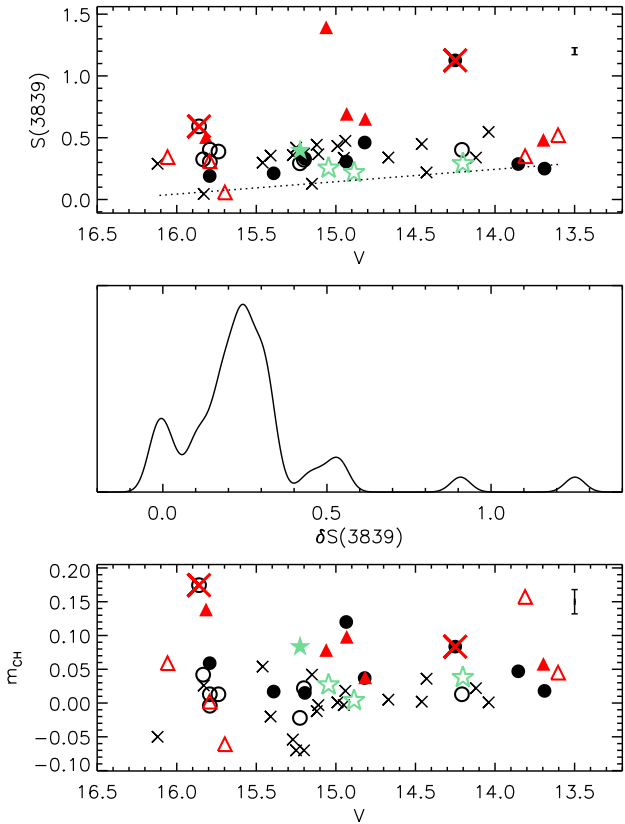}
      \caption{The \scn\ CN index versus $V$ magnitude (upper), the
distribution of CN excess $\delta$ \scn\ (middle) and the \mch\ CH index versus
$V$ magnitude (lower). (We exclude stars 82, 184, 648 and 707 since they are
not on the RGB.) The program stars are shown as black circles (canonical RGB),
red triangles (metal-poor anomalous RGB) and aqua stars (metal-rich anomalous
RGB). Filled symbols are proper-motion members according to \citet{cudworth87}.
The crosses are stars from \citet{smith90}, and the two CH objects are
indicated by large red crosses. A representative error bar is shown in the top
and bottom panels. 
      \label{fig:cnch} }
\end{figure}

In the upper panel of Figure \ref{fig:cnch}, we plot the \scn\ index against
$V$ mag.  In this figure, we include the data from \citet{smith90} and exclude
the UV-bright (NR 184), HB (NR 648 and NR 707) and AGB (NR 82) stars.  As
discussed in Section 3.1, our measurements are on the same scale as
\citet{smith90}.  The middle panel shows the generalised histogram of the \scn\
residuals, $\delta$\scn, measured with respect to the same baseline as in
\citet{smith90}, namely $S_0(3839) = -0.1V + 1.644$.  The generalised histogram
was produced using a Gaussian kernel with a FWHM of 0.03. We note that while
\citet{smith90} identified a particularly CN rich star (HI-240, \scn\ = 1.110),
our sample includes an even more extreme example, NR 358 with \scn\ = 1.394. In
the following subsection, however, we note that NR 358 (not observed at high
resolution) has a CMD location inconsistent with cluster membership given the
metallicity of this star assuming no significant age spread in the cluster. 

In the lower panel of Figure \ref{fig:cnch}, we plot the \mch\ index against
$V$ mag. Consideration of the measurement errors would indicate a genuine
spread in the \mch\ index within this cluster. In addition to the two CH stars
identified by \citet{smith90}, there are three stars with \mch~$>$~0.1, NR~81,
NR~299 and NR~1204. Given the metallicity of NR~1204, the CMD location is
inconsistent with cluster membership (i.e., we use the same argument as for
NR~358 above that will be described in the following subsection). We have no
reason to suspect non-membership for the other two stars with strong \mch\
indices, NR~81 and NR~299. There is no obvious anti-correlation between the
\scn\ and \mch\ indices. Indeed, the two CH stars from \citet{smith90} also
exhibit large \scn\ indices. {\it The first key result is that we confirm the
presence of unusually CN and/or CH strong stars in M2.} 

\subsection{Calcium triplet and high-resolution metallicities} 

Based on the iron abundances derived from the high dispersion spectra, it is
clear that the anomalous RGB stars have higher [Fe/H] values than those for the
normal RGB stars.  In particular, the six normal RGB stars in Table
\ref{tab:param} have a mean iron abundance of $\langle$[Fe/H]$\rangle$ =
$-$1.67 $\pm$ 0.02 ($\sigma$ = 0.04).  The eight anomalous stars separate into
two metallicity groups (and in the following subsection we shall see that the
two groups exhibit distinct [X/Fe] ratios). The more metal-poor group of
anomalous RGB stars includes four objects (NR 38, NR 47, NR 77 and NR 81) and
has $\langle$[Fe/H]$\rangle$ = $-$1.51 $\pm$ 0.04 ($\sigma$ = 0.09) dex.  The
more metal-rich group of anomalous RGB stars consists of four objects (NR 132,
NR 207, NR 254 and NR 378) and has $\langle$[Fe/H]$\rangle$ = $-$1.03 $\pm$
0.03 ($\sigma$ = 0.06) dex. When defined in this way, each of the three groups
of stars (canonical RGB, metal-poor anomalous RGB and metal-rich anomalous RGB)
likely have metallicities consistent with a single value, i.e., the dispersion
in [Fe/H] for a given group can probably be explained entirely by the
measurement uncertainties.  We now turn to the \caii\ triplet spectra to
investigate the presence of a metallicity dispersion in this cluster. 

\begin{figure}
\centering
      \includegraphics[width=0.99\hsize]{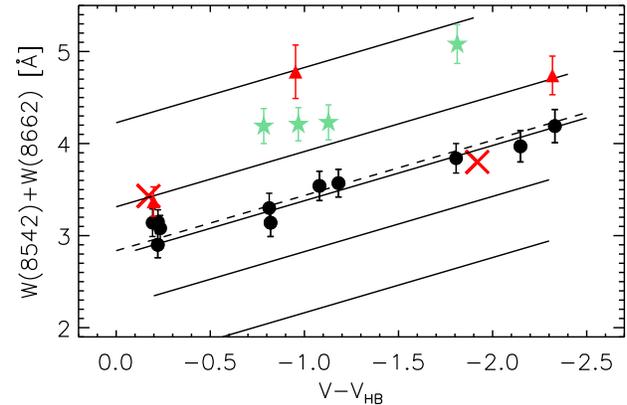}
      \caption{The sum of the EWs of the \caii\ triplet lines at 8542\,\AA\ and
8662\,\AA\ are plotted against magnitude difference from the horizontal branch
$V-V_{HB}$.  M2 stars lying on the ``normal'' RGB are shown as black circles
while stars from the ``anomalous'' RGB are plotted as red triangles or aqua
star symbols. The two CH-stars identified by Smith \& Mateo (1990) are shown as
red crosses. The solid lines are the relations between summed EWs and
$V-V_{HB}$ for the standard clusters.  In order of increasing summed EW the
standard clusters are NGC~7099 ([Fe/H] = $-$2.27), NGC~2298 ($-$1.96), NGC~1904
($-$1.58), NGC~288 ($-$1.32) and 47~Tuc ($-$0.76).  The dashed line is a fit of
a line with slope $-$0.60 \AA/mag to the M2 normal RGB stars. 
      \label{fig_CaT}}
\end{figure}

In Figure \ref{fig_CaT} we plot the sum of the EWs of the two
stronger \caii\ triplet lines against the magnitude difference from the
horizontal branch, $V-V_{HB}$ for the M2 stars observed at this wavelength
setting with AAOmega.  Here the $y$ magnitudes were assumed to be equivalent to
$V$ and the value of $V_{HB}$ was taken from Harris (1996).  ``Normal'' RGB
stars are plotted as black circles while the triangle and star symbols show the
location of stars from the anomalous RGB\@.  The two CH-stars identified by
Smith \& Mateo (1990) are shown as red crosses. 

In order to calibrate the line strengths in terms of [Fe/H] we have
made use of similar observations of red giants in ``standard'' globular
clusters that have well established abundances.  The clusters are NGC~7099
(M30), NGC~2298, NGC~1904, NGC~288 and 47~Tuc.  The observations for these
clusters were obtained with AAOmega during the same observing run as that for
the M2 \caii\ triplet spectra, using an identical instrumental setup.  The
standard cluster stars observed were chosen using the photometry lists made
publicly available by Peter Stetson at the Canadian Astronomy Data
Centre\footnote{www3.cadc-ccda.hia-iha.nrc-cnrc.gc.ca/community/STETSON/standards/}.
A similar analysis to that described here for M2 led to the identification of
RGB cluster members from the observations.  The \caii\ triplet line strengths of
these stars were then measured using the same procedure as for the M2 stars
described in Section 3.1.
The numbers of confirmed RGB cluster members ranged from 8 and 10 in NGC~2298
and NGC~7099 to 33 and 46 in NGC~288 and 47~Tuc.  In each cluster the RGB stars
covered at least two magnitudes in $V-V_{HB}$ at luminosities exceeding
$V-V_{HB}$ $\approx$ 0.0 mag, and we adopted $V_{HB}$ from
\citet{harris96}. A slope of $-$0.60 $\pm$ 0.01 \AA/mag was found
to fit consistently each set of  cluster data.  This value is similar to that
found in other \caii\ triplet studies: for example, \citet{saviane12} find a
value for the slope of $-$0.627 while the original study of \citet{AD91} found
a slope of $-$0.619 \AA/mag. 

Adopting W$^\prime$ as the value of the summed EW at $V-V_{HB}$ = 0 with the
adopted slope of $-$0.60 \AA/mag, and [Fe/H] abundances from \citet{Ca09} for
the standard clusters, then yields a very well defined linear relationship
between W$^\prime$ and [Fe/H]: [Fe/H] = 0.590~W$^\prime$ $-$ 3.253 dex.  The
RMS about the fitted relation is only 0.02 dex, indicating excellent
consistency between the \caii\ triplet line strength measurements for these
clusters and the \citet{Ca09} [Fe/H] abundances.  The relation is valid for the
abundances encompassed by the standard clusters, i.e., from [Fe/H] $\approx$
$-$2.3 to [Fe/H] $\approx$ $-$0.7 dex. 

Returning now to Figure \ref{fig_CaT}, we note that the normal M2 RGB stars
cluster tightly around the fitted line of slope $-$0.60 \AA/mag, shown as the
dashed line.  In particular, there is no evidence for any intrinsic dispersion
in [Fe/H] values from the \caii\ triplet line strengths of these stars.  The
[Fe/H] abundance derived from the mean W$^\prime$ value is [Fe/H]$_{\rm CaT}$ =
$-$1.58 $\pm$ 0.08 dex, where the error includes the RMS deviation about the
fitted line for the 11 normal RGB stars and the (minor) calibration
uncertainty.  This value of [Fe/H] is somewhat higher than the value listed in
\citet{Ca09} for M2, [Fe/H] = $-$1.66 $\pm$ 0.07, and in the latest version of
the Harris (1996) catalogue ([Fe/H] = $-$1.65).  Both of these values stem from
the measurement of \caii\ triplet line strengths in an integrated spectrum of
M2 obtained by \citet{AZ88}. The value is also somewhat higher than the mean
abundance, $-$1.67 $\pm$ 0.02 (std.\ error of mean), of the six normal RGB
stars observed at high dispersion. 

Nevertheless, there is good agreement between the [Fe/H] values derived from
the \caii\ line strengths and from high dispersion analysis for the three normal
RGB stars in common (NR~76, 99 and 124). For these three stars, the mean
difference in [Fe/H], in the sense of the high dispersion values minus the
\caii\ values, is $-$0.03 $\pm$ 0.01 dex ($\sigma$ = 0.02).  This consistency
also applies to the 5 anomalous RGB stars (NR~47, 132, 207, 254 and 378) in
common between the two datasets.  Here the mean difference is 0.00 $\pm$ 0.05
dex ($\sigma$ = 0.11) suggesting we can combine the [Fe/H] determinations for
the anomalous RGB stars into a single sample.  There are then 10 anomalous RGB
star [Fe/H] determinations, eight from the high dispersion analysis, seven from
the \caii\ triplet spectroscopy with five stars in common.  For the latter
stars the [Fe/H] values have been averaged, weighted by the uncertainties.  We
assume for the present that all the stars are cluster members. 

The mean abundance of the anomalous RGB stars is the $\langle$[Fe/H]$\rangle$ =
$-$1.29 $\pm$ 0.09, considerably more metal-rich than that for the normal RGB
stars, and with a substantial dispersion of 0.28 dex.  The [Fe/H] range shown
by the anomalous RGB stars is $\sim$0.8 dex indicating that there is a
substantial intrinsic iron abundance spread present.  Moreover, the value of
the mean abundance and the size of the intrinsic abundance spread do not change
significantly even if the sample is restricted to the four anomalous RGB stars
with 99\% membership probabilities.  Further, although the sample is not large,
the anomalous RGB stars appear to fall into two distinct metallicity groups,
each containing 5 objects.  The first, consisting of stars NR~38, 47, 77, 81
and 1204, has a mean abundance of $\langle$[Fe/H]$_{\rm CaT}\rangle$ = $-$1.47
$\pm$ 0.05 ($\sigma$ = 0.11). For the four stars in this group with high
dispersion spectra, the mean abundance is $\langle$[Fe/H]$\rangle$ = $-$1.51
$\pm$ 0.04 ($\sigma$ = 0.09). Similarly, for the second group of stars, NR~132,
207, 254, 358 and 378, the mean abundance is $\langle$[Fe/H]$_{\rm CaT}\rangle$
= $-$0.98 $\pm$ 0.06 ($\sigma$ = 0.13) dex and for the four stars with high
dispersion spectra, the mean abundance is $\langle$[Fe/H]$\rangle$ = $-$1.03
$\pm$ 0.03 ($\sigma$ = 0.06). Within each group the intrinsic abundance
dispersion is notably smaller than for the full sample, and these two groups
mirror those identified by our high resolution spectroscopic analysis. 

We note in passing that we have not included the two CH stars in the above
discussion.  Nevertheless, the \caii\ triplet spectra of these two objects
appear very similar to those of the other M2 stars observed.  The measured line
strengths imply abundances of [Fe/H]$_{\rm CaT}$ = $-$1.69 $\pm$ 0.11 for
HI-240 and [Fe/H]$_{\rm CaT}$ = $-$1.29 $\pm$ 0.12 for HI-451.  The former is
consistent with that for the normal RGB stars as well that of anomalous RGB
stars such as NR~38 ([Fe/H]$_{\rm CaT}$ = $-$1.61 $\pm$ 0.05).  The latter is
similar to those for the anomalous RGB stars NR~207 ([Fe/H]$_{\rm CaT}$ =
$-$1.11 $\pm$ 0.07) and NR~1204 ([Fe/H]$_{\rm CaT}$ = $-$1.34 $\pm$ 0.09). 

In the above discussion we have implicitly assumed that the stars observed are
all members of M2, deriving abundances under that assumption.  There seems no
reason to doubt the membership of any of the stars in the normal RGB samples.
There is, however, a consistency check that we can apply to further investigate
the membership status of the anomalous RGB stars.  The check is as follows:
given the reasonable assumption that the age range in M2 is small ($\lesssim$2
Gyr, \citealt{piotto12}), stars that are M2 members with higher [Fe/H]
abundances should lie to the red of normal RGB stars at the same magnitude in
the CMD by an amount that depends on the excess in [Fe/H] above that for the
normal RGB stars.  Ideally such an investigation would use, for example, an
($I$, $V-I$) CMD to minimise the potential influence of molecular bands on the
photometry at bluer wavelengths.  However, such photometry is not available for
most of the anomalous RGB stars.  We have therefore used a ($V$, $B-V$) CMD
based on the M2 photometry given in Stetson's Photometric Standard Star fields
available from the Canadian Astronomy Data Centre.  The M2 normal RGB is
well-defined in this data set.  We then plotted the stars observed
spectroscopically in the CMD using either Stetson's photometry where available
or by generating $V$ and $(B-V)$ values from the \citet{grundahl99} $y$ and
$(b-y)$ photometry.  Here we have $y$ = $V$ and $(B-V)$ = 1.64 $(b-y)$ with the
latter relation determined from 11 stars in common between Stetson's photometry
list and the stars observed at the \caii\ triplet.  The RMS deviation about the
relation is only 0.009 mag. 

We then use isochrones for metallicities of [Fe/H] = $-$1.65, $-$1.25 and
$-$0.85 dex, [$\alpha$/Fe] = +0.4 and an age of 13 Gyr from the Dartmouth
isochrone set \citep{Dotter08} to provide an indication of the colour shift
expected for the metallicities of the anomalous RGB stars.  We adopt values
from the current on-line version of the Harris (1996) database for the
reddening and distance modulus of M2 and with these parameters the [Fe/H] =
$-$1.65 theoretical  RGB is an acceptable representation of the normal RGB
stars in the CMD.   

Specifically, for each anomalous RGB star, we have interpolated in the
isochrones at the $V$ magnitude of the star to determine the $(B-V)$ that
corresponds to the [Fe/H] value.  This colour, and its uncertainty derived from
the uncertainty in the [Fe/H] value, is then compared with the observed $(B-V)$
value.  Stars NR~38, 47, 77 and 81 have predicted colours that agree well with
the observed colours on the metal-poor anomalous RGB: the mean difference
(observed -- predicted) is 0.00 $\pm$ 0.03 with, in each case, the predicted
colour lying within 2$\sigma$ of the observed colour.  We conclude therefore
that all four of these stars are likely to be members of the cluster: one
(NR~47) has a 99\% membership probability from Cudworth \& Rauscher (1987)
while the others are not classified.  Conversely, with this approach it seems
probable that stars NR~132, 358, 378 and 1204 are not members of the cluster.
Here the colour differences on the metal-rich anomalous RGB are --0.18 $\pm$
0.04, --0.29 $\pm$ 0.04, --0.15 $\pm$ 0.03, and --0.08 $\pm$ 0.02,
respectively; i.e., in each case the location of the star in the CMD is at
least 3.5$\sigma$ bluer than predicted for the star's metallicity.   The
observed colours can only be reproduced if the age of the stars is at least 6
Gyr younger than the 13 Gyr assumed, which seems unlikely, although as we have
already noted in passing, this argument only considers metallicity and that
other elements (He, C, N, O and $\alpha$ elements) can also affect the $B-V$
colour.  We note further that our classification contrasts with the fact that
three of these stars (NR~358, 378 and 1204) have 99\% membership probabilities
in the Cudworth \& Rauscher (1987) study.\footnote{As discussed in
\citet{cudworth87}, the proper motions membership probabilities are based on a
relative system with the zero point set by the mean of all the measurements.
Since the M2 sample is dominated by cluster members (see Table II of Cudworth
\& Rauscher 1987) whose absolute proper motions will be small given the large
distance, any relatively distant field star, as distinct from nearby dwarfs,
will likely also have a small proper motion and therefore potentially be
assigned an erroneous high membership probability.} For stars NR~207 and NR~254
the comparison suggests that these stars may also be non-members: both lie in
the CMD 0.10 $\pm$ 0.03 mag bluer than the predicted colour.  Neither has a
classification in the \citet{cudworth87} study.  For these stars we will need
to rely on the similarity of their chemical abundance distributions with those
of the cluster members, or with the non-members, for the membership
classification. 

We conclude therefore that at least some of the anomalous RGB stars, in
particular the stars NR~38, NR~47, NR~77 and NR~81 (and perhaps also NR~207 and
NR~254) are likely bona-fide members of M2. {\it If this is indeed the case
then the second key result we find is that M2 joins other clusters like M22
\citep{dacosta09,marino09,marino11}, M54 \citep{carretta10,saviane12} and NGC
5824 \citep{dacosta14} in possessing a modest intrinsic [Fe/H] range: M2 has
member stars with [Fe/H] values up to 0.25 dex above that for the majority of
cluster members, perhaps up to 0.7 dex depending on the membership status of
the metal-rich anomalous RGB stars.} 

To investigate the likelihood of observing field stars in the vicinity of M2
with stellar parameters (\teff, \logg\ and [Fe/H]) similar to that of the four
metal-rich anomalous RGB stars observed at high resolution, we make use of the
Trilegal Galactic model \citep{girardi05}.  First, we consider all stars within
a one degree square field centered on M2.  Secondly, we restricted the sample
to lie in the same region in the $v$ versus $u-y$ CMD from which we selected
the anomalous RGB stars.  We find 17285 such stars in the Trilegal model.
Thirdly, of these 17285 stars, we counted the number that satisfied the
following constraints: ($i$) $-$25 $\le$ RV $\le$ $+$25 \kms\ and ($ii$) $-$1.2
$\le$ [Fe/H] $\le$ $-$0.8 dex. And finally, we counted the numbers of stars
that lay in a particular region in the \teff-\logg\ plane, specifically, the
area is bounded at the left edge by the line from (\teff,\logg) = (5000,2.0) to
(\teff,\logg) = (4300,0.0), at the right edge by the line from (\teff,\logg) =
(4400,2.0) to (\teff,\logg) = (3700,0.0) both with 0.0 $\le$ \logg\ $\le$ 2.0.
We found 46 stars in the Trilegal model that satisfied all criteria and
therefore estimate that given a sample of stars occupying similar locations in
the $v$ versus $u-y$ CMD as the program stars, the probability of observing a
field star with stellar parameters and a radial velocity consistent with the
metal-rich population is roughly 0.3\%. We reach similar conclusions when using
the Besan\c{c}on model \citep{robin03}.  Accurate proper-motion and parallax
measurements from GAIA will establish cluster membership, or otherwise, for the
M2 stars. 

Given the strong bias towards anomalous RGB stars in the samples selected for
observation here, we have little constraint on the form of the iron abundance
distribution function other than noting that the normal RGB population is
dominant and the anomalous RGB is not prominent (e.g., \citealt{lardo12}). In
this context, the anomalous fainter subgiant branch contains only a small
fraction of stars, $\sim$4\%, relative to the dominant brighter subgiant branch
\citep{piotto12}. An unbiased sample of RGB stars is needed to constrain the
iron abundance distribution and allow comparison with those of other clusters.
We now examine the element-to-iron abundance ratios from the high dispersion
spectra of the normal and anomalous RGB stars. 

\subsection{Chemical Abundance Ratios} 

\begin{figure}
\centering
      \includegraphics[width=0.99\hsize]{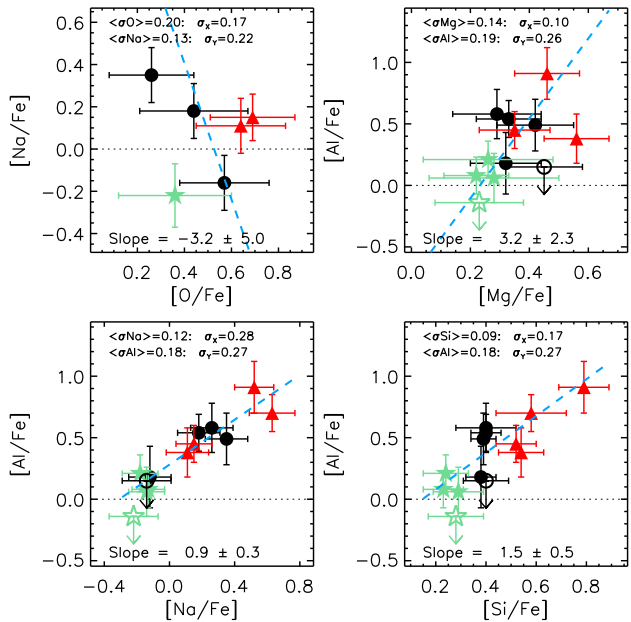}
      \caption{Abundance ratios for combinations of the light elements (O, Na,
Mg, Al and Si) for the stars observed at high spectral resolution.  The black
points are stars on the canonical RGB while the red and aqua points are stars
on the anomalous RGB. The aqua points are the unusually metal-rich objects.
Open symbols reflect upper limits. The dashed blue line is the linear fit to
the data (slope and error are included).  The average error
($<\sigma$[X/Fe]$>$) and dispersion ($\sigma$) in the x-direction and
y-direction are included. 
      \label{fig:onamgalsi} }
\end{figure}

In Figure \ref{fig:onamgalsi}, we plot combinations of the light elements (O,
Na, Mg, Al and Si) against one another. M2 exhibits star-to-star abundance
variations of the light elements along with the usual correlations and
anti-correlations between these elements found in globular clusters (e.g., see
reviews by \citealt{smith87}; \citealt{kraft94};
\citealt{gratton04,gratton12}). In particular, we note that the observed
dispersions in [X/Fe] for Na, Al and Si are considerably larger than the
average measurement uncertainties indicating genuine abundance spreads.  The
six canonical RGB stars (black circles in Figure \ref{fig:onamgalsi}) clearly
exhibit abundance dispersions for Na and Al as well as a correlation between
these elements. The four metal-poor anomalous RGB stars (red triangles in
Figure \ref{fig:onamgalsi}) also exhibit these abundance patterns and this
would suggest that they are cluster members. The four metal-rich anomalous RGB
stars (aqua star symbols in Figure \ref{fig:onamgalsi}) do not exhibit
abundance variations for Na and Al. On the other hand, Si does not usually
exhibit a star-to-star abundance variation within a given cluster, with a
handful of exceptions including NGC 6752 \citep{yong05} and NGC 4833
\citep{carretta14}. For O and Mg, there is no compelling evidence for an
abundance dispersion within our sample. 

\begin{figure}
\centering
      \includegraphics[width=0.99\hsize]{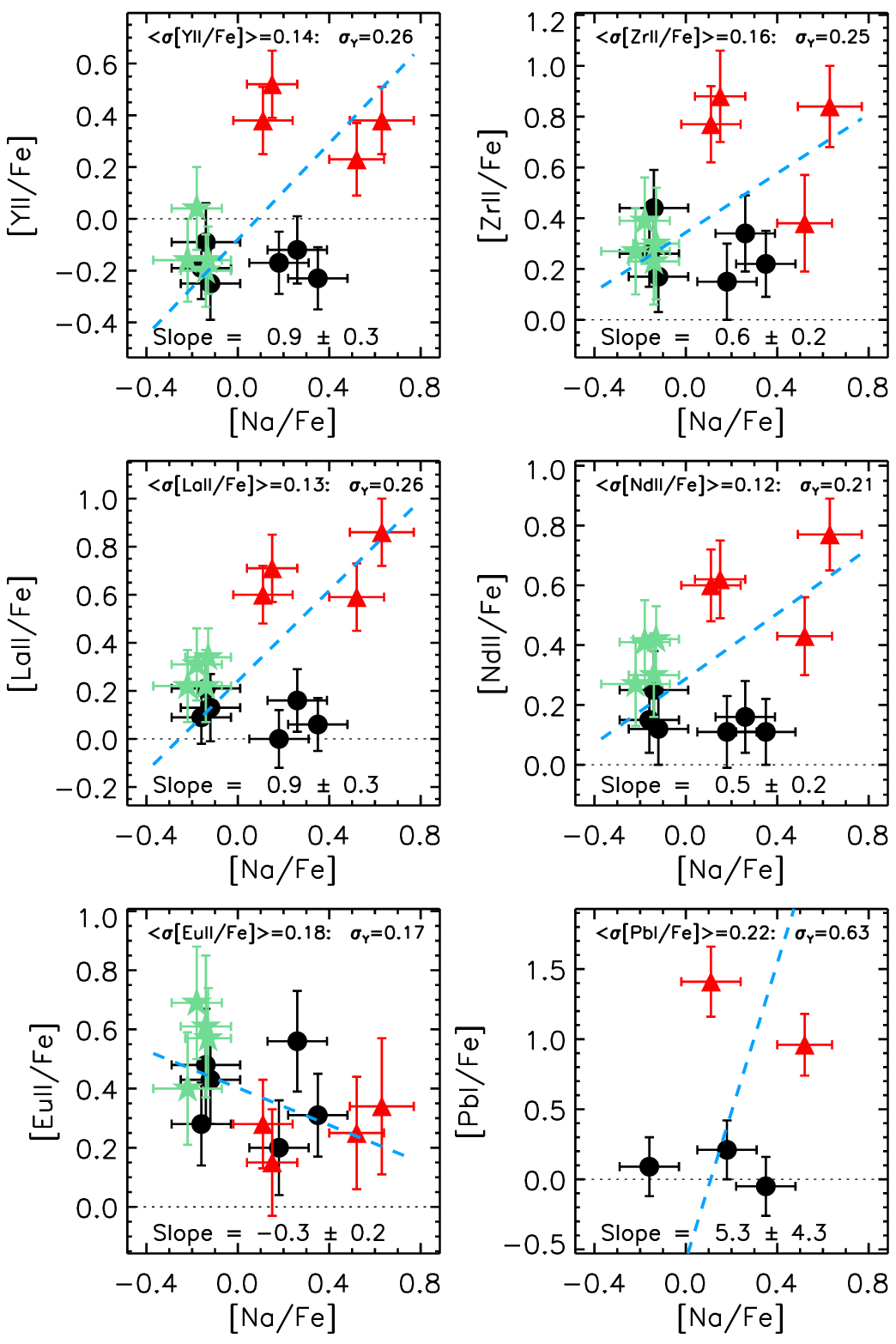}
      \caption{Same as Figure \ref{fig:onamgalsi} but for neutron-capture elements
versus [Na/Fe]. 
      \label{fig:sna} }
\end{figure}

Next, in Figure \ref{fig:sna}, we plot [X/Fe] versus [Na/Fe] for six
neutron-capture species (Y, Zr, La, Nd, Eu and Pb). While there is no evidence
for any significant trend between [X/Fe] versus [Na/Fe], it is clear that the
four stars on the anomalous RGB with [Fe/H] $\approx$ $-$1.5 
exhibit large overabundances of the $s$-process elements with respect to the
six stars on the canonical RGB. Such a result is not unexpected given the
clear star-to-star line strength differences for neutron-capture elements seen
in Figures \ref{fig:spec1s} to \ref{fig:spec2m}. Confirmation of the presence
of a large spread in neutron-capture element abundances can be obtained by 
noting that the observed dispersion exceeds the average measurement uncertainty. 
{\it The third key result is that we identify an intrinsic abundance dispersion
for the neutron-capture elements in M2 thereby verifying and extending the
results of \citet{lardo13}.} M2 joins the small, but growing, group of globular
clusters that exhibit abundance variations for the neutron-capture elements as
well as iron abundance dispersions. These clusters include $\omega$
Cen, M22 and NGC 1851
\citep{norris95,smith00,yong081851,marino09,marino11,villanova09,johnson10,carretta11,dorazi11,roederer11}.
Additionally, there are other globular clusters with a dispersion in
neutron-capture element abundances, but no obvious iron abundance dispersion
including M15 \citep{sneden97,sneden00,otsuki06,sobeck11,worley13} and NGC 362
\citep{carretta13}. 

\begin{figure*}
\centering
      \includegraphics[width=0.65\hsize]{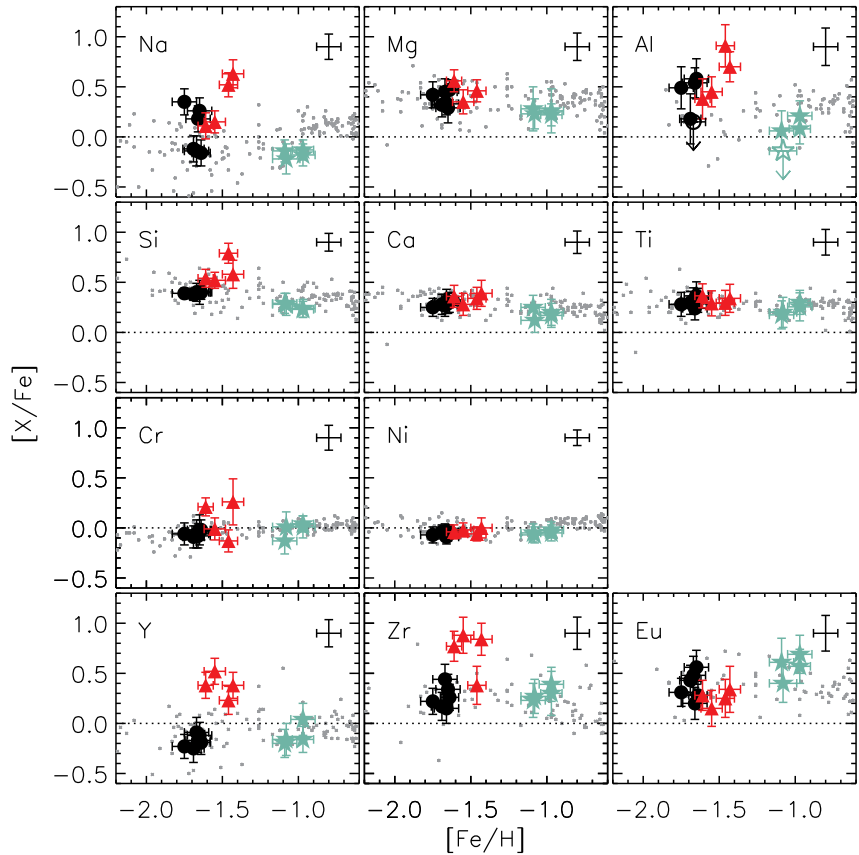}
      \caption{Abundance ratios [X/Fe] versus [Fe/H] for the program stars
observed at high spectral resolution. The colours are the same as in Figure
\ref{fig:onamgalsi}. The grey symbols are field halo stars taken from
\citet{fulbright00}. 
      \label{fig:abundfield} }
\end{figure*}

In Figure \ref{fig:abundfield} we plot [X/Fe] versus [Fe/H] for the program stars
and field stars from \citet{fulbright00}. Here one sees that the six M2 giants
on the canonical RGB appear to follow the trends exhibited by field halo stars
(although we recognise that there may be systematic abundance differences
between this analysis and that of \citealt{fulbright00}).  Similarly, in this
figure the four $s$-process rich stars with [Fe/H] $\approx$ $-$1.5 have [X/Fe]
ratios (excluding Y and Zr) consistent with field stars at the same
metallicity. For both sets of stars, Na and Al may exhibit higher abundance
ratios compared to field stars at the same metallicity.  For the four
metal-rich stars with [Fe/H] $\approx$ $-$1.0, the abundance ratios for all
elements included in this figure are consistent with field stars of comparable
metallicity. 

\begin{figure*}
\centering
      \includegraphics[width=0.9\hsize]{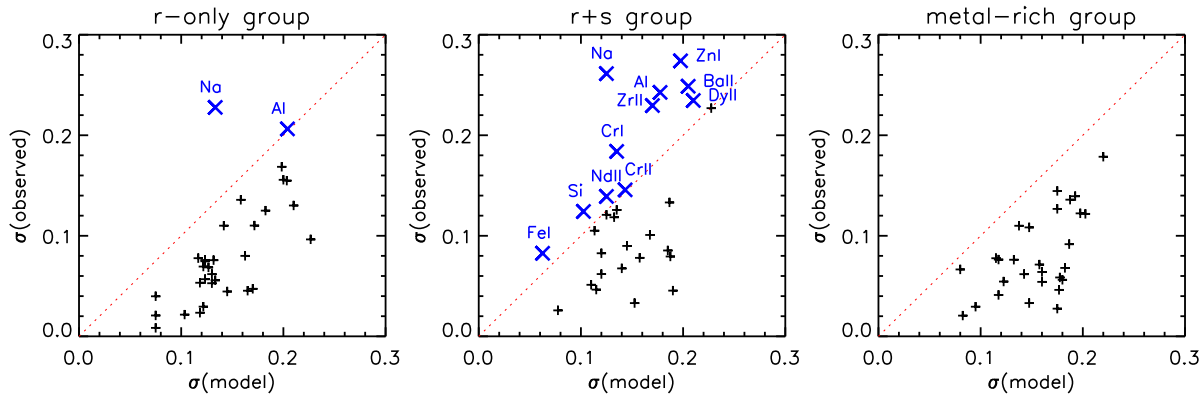}
      \caption{Measured abundance dispersion $\sigma$(observed) versus average
measurement error $\sigma$(model) for the $r$-only (left panel), $r+s$
(middle panel) and metal-rich (right panel) groups. The dotted red line is
the 1:1 relation. Elements which fall on or above the 1:1 relation
are plotted as large blue crosses and the species names are written.
      \label{fig:err} }
\end{figure*}

In Figure \ref{fig:err}, we compare the average measurement
errors\footnote{For a given element in a set of stars, the ``average
measurement error'' is the average of the Total Error presented in Table
\ref{tab:abun}.} 
with the
observed dispersion in [X/Fe] ratios for the three groups of stars: (1) the six
canonical RGB objects (NR~37, 58, 60, 76, 99 and 124), (2) the four $s$-process
rich anomalous RGB stars (NR~38, 47, 77 and 81) and (3) the four metal-rich
anomalous RGB stars (NR~132, 207, 254, 378).  For the second group, all are
likely members based on our analysis in the previous subsection whereas for the
third group, their membership is questionable based on the analysis presented
in Section 4.4, although \hst\ photometry suggests that these stars may indeed
be members (Milone et al.\ in preparation).  For reasons that will become
clearer in the following section, we refer to the three groups as the
$r$-process only group (``$r$-only''), the $r$- + $s$-process group (``$r+s$'')
and the ``metal-rich'' groups, respectively. For the purposes of this exercise,
we assumed that the [Al/Fe] limits are detections, and therefore the observed
dispersion for [Al/Fe] in the $r$-only group is effectively a lower limit. In
general, there is a suggestion that the abundance errors are overestimated as
the majority of elements lie below the 1:1 relation.  For the $r$-only group,
only Na (and perhaps Al) exhibits an abundance dispersion that significantly
exceeds the average measurement error.  For the $r+s$ group, a handful of
elements including Na, Al, Cr, Zn, Zr and Ba exhibit abundance dispersions that
exceed the average measurement error. For the metal-rich group, all elements
exhibit abundance dispersions that are consistent with the expected dispersion
given the average measurement error. That said, it is important to emphasise
that for most elements, there is no evidence for an intrinsic abundance
dispersion within a given group of stars.  That is, with the exception of a few
elements in the ``$r$-only'' and ``$r+s$'' groups, the dispersion in [X/Fe] is
consistent with the measurement error. 

\section{DISCUSSION} 

The aim of this discussion is to examine the nature of M2 in light of the
chemical abundance ratios with an emphasis on the neutron-capture elements
(Sections 5.2 and 5.3). In Table \ref{tab:meanabun}, we present the average
abundance ratios and dispersions for $\log\epsilon$ (X) and [X/Fe] for the
$r$-only, $r+s$ and metal-rich groups of stars. 

\begin{table*}
 \centering
 \begin{minipage}{180mm}
  \caption{Mean chemical abundances for the three groups of stars.}
  \label{tab:meanabun} 
  \begin{tabular}{@{}lrcrccrcrccrcrc@{}}
  \hline
	Species & 
	$<\log\epsilon>$ &
	$\sigma$\footnote{These values are the standard deviation.} &
	$<$[X/Fe]$>$ & 
	$\sigma$ & 
        &
	$<\log\epsilon>$ &
	$\sigma$ &
	$<$[X/Fe]$>$ & 
	$\sigma$ & 
	& 
	$<\log\epsilon>$ &
	$\sigma$ &
	$<$[X/Fe]$>$ & 
	$\sigma$ \\ 
	\cline{2-5}  
	\cline{7-10} 
	\cline{12-15}
	&
        \multicolumn{4}{c}{Six stars in the $r$-only group} &
	&
        \multicolumn{4}{c}{Four stars in the $r+s$ group} &
	& 
        \multicolumn{4}{c}{Four stars in the ``metal-rich'' group}\\
\hline 
 \oi  &    7.43 &    0.21 &    0.42 &    0.16 &  &    7.77 &    0.08 &    0.66 &    0.04 &  &    7.97 &  \ldots &    0.36 &  \ldots \\
\nai  &    4.63 &    0.21 &    0.06 &    0.23 &  &    5.07 &    0.34 &    0.35 &    0.26 &  &    5.04 &    0.09 & $-$0.17 &    0.04 \\
\mgi  &    6.30 &    0.08 &    0.38 &    0.08 &  &    6.51 &    0.10 &    0.46 &    0.11 &  &    6.82 &    0.06 &    0.25 &    0.03 \\
\ali  &    5.22 &    0.20 &    0.45 &    0.18 &  &    5.55 &    0.32 &    0.61 &    0.24 &  &    5.56 &    0.13 &    0.12 &    0.08 \\
\sii  &    6.23 &    0.05 &    0.40 &    0.01 &  &    6.61 &    0.19 &    0.61 &    0.12 &  &    6.74 &    0.03 &    0.26 &    0.03 \\
\cai  &    4.95 &    0.06 &    0.28 &    0.02 &  &    5.17 &    0.11 &    0.34 &    0.05 &  &    5.50 &    0.09 &    0.19 &    0.05 \\
\scii &    1.45 &    0.09 & $-$0.03 &    0.06 &  &    1.58 &    0.10 & $-$0.06 &    0.08 &  &    1.99 &    0.12 & $-$0.13 &    0.06 \\
\tii  &    3.45 &    0.05 &    0.17 &    0.02 &  &    3.75 &    0.13 &    0.32 &    0.05 &  &    4.07 &    0.14 &    0.14 &    0.08 \\
\tiii &    3.70 &    0.09 &    0.43 &    0.07 &  &    3.76 &    0.06 &    0.33 &    0.09 &  &    4.24 &    0.11 &    0.32 &    0.05 \\
\cri  &    3.91 &    0.06 & $-$0.06 &    0.03 &  &    4.21 &    0.19 &    0.08 &    0.18 &  &    4.60 &    0.13 & $-$0.02 &    0.08 \\
\crii &    4.11 &    0.10 &    0.14 &    0.08 &  &    4.22 &    0.07 &    0.12 &    0.15 &  &    4.74 &    0.15 &    0.11 &    0.09 \\
\mni  &    3.34 &    0.07 & $-$0.41 &    0.05 &  &    3.51 &    0.03 & $-$0.41 &    0.06 &  &    4.04 &    0.12 & $-$0.36 &    0.07 \\
\fei\footnote{This is [\fei/H] or [\feii/H], not $<$[X/Fe]$>$.}  &    5.82 &    0.04 & $-$1.68 &    0.04 &  &    5.99 &    0.08 & $-$1.51 &    0.08 &  &    6.47 &    0.07 & $-$1.03 &    0.07 \\
\feii$^b$ &    5.83 &    0.06 & $-$1.66 &    0.06 &  &    5.99 &    0.10 & $-$1.51 &    0.10 &  &    6.48 &    0.06 & $-$1.02 &    0.06 \\
\coi  &    3.19 &    0.02 & $-$0.13 &    0.05 &  &    3.46 &    0.10 & $-$0.02 &    0.07 &  &    3.87 &    0.11 & $-$0.11 &    0.05 \\
\nii  &    4.49 &    0.05 & $-$0.05 &    0.02 &  &    4.68 &    0.10 & $-$0.03 &    0.03 &  &    5.14 &    0.08 & $-$0.05 &    0.02 \\
\cui  &    1.84 &    0.07 & $-$0.68 &    0.05 &  &    2.28 &    0.13 & $-$0.40 &    0.09 &  &    2.63 &    0.19 & $-$0.53 &    0.13 \\
\zni  &    2.93 &    0.13 &    0.04 &    0.11 &  &    3.13 &    0.29 &    0.08 &    0.27 &  &    3.56 &    0.10 &    0.02 &    0.14 \\
\sri  &    0.63 &    0.07 & $-$0.56 &    0.10 &  &    1.31 &    0.20 & $-$0.04 &    0.23 &  &    1.22 &    0.22 & $-$0.62 &    0.18 \\
\yii  &    0.36 &    0.09 & $-$0.18 &    0.06 &  &    1.07 &    0.11 &    0.38 &    0.12 &  &    1.06 &    0.16 & $-$0.12 &    0.11 \\
\zri  &    0.83 &    0.15 & $-$0.08 &    0.17 &  &    1.62 &    0.07 &    0.56 &    0.05 &  &    1.56 &    0.12 &    0.01 &    0.06 \\
\zrii &    1.17 &    0.13 &    0.26 &    0.11 &  &    1.78 &    0.22 &    0.72 &    0.23 &  &    1.85 &    0.13 &    0.30 &    0.07 \\
\moi  &    0.10 &  \ldots & $-$0.03 &  \ldots &  &  \ldots &  \ldots &  \ldots &  \ldots &  &  \ldots &  \ldots &  \ldots &  \ldots \\
\baii &    0.69 &    0.17 &    0.19 &    0.15 &  &    1.59 &    0.28 &    0.92 &    0.25 &  &    1.45 &    0.16 &    0.30 &    0.12 \\
\laii & $-$0.47 &    0.09 &    0.11 &    0.07 &  &    0.28 &    0.18 &    0.69 &    0.13 &  &    0.34 &    0.13 &    0.27 &    0.06 \\
\ceii & $-$0.10 &    0.09 & $-$0.01 &    0.07 &  &    0.57 &    0.14 &    0.50 &    0.12 &  &    0.72 &    0.09 &    0.16 &    0.03 \\
\prii & $-$0.88 &    0.07 &    0.08 &    0.04 &  & $-$0.30 &    0.08 &    0.49 &    0.03 &  & $-$0.12 &    0.20 &    0.19 &    0.14 \\
\ndii & $-$0.10 &    0.08 &    0.15 &    0.05 &  &    0.51 &    0.17 &    0.61 &    0.14 &  &    0.74 &    0.14 &    0.35 &    0.08 \\
\smii & $-$0.45 &    0.10 &    0.26 &    0.08 &  & $-$0.12 &    0.10 &    0.43 &    0.08 &  &    0.44 &    0.17 &    0.50 &    0.11 \\
\euii & $-$0.78 &    0.15 &    0.38 &    0.14 &  & $-$0.74 &    0.14 &    0.25 &    0.08 &  &    0.06 &    0.17 &    0.57 &    0.12 \\
\gdii & $-$0.28 &    0.09 &    0.33 &    0.08 &  & $-$0.05 &    0.06 &    0.43 &    0.13 &  &    0.80 &  \ldots &    0.70 &  \ldots \\
\tbii & $-$1.13 &    0.07 &    0.27 &    0.14 &  &  \ldots &  \ldots &  \ldots &  \ldots &  &  \ldots &  \ldots &  \ldots &  \ldots \\
\dyii & $-$0.25 &    0.14 &    0.33 &    0.12 &  &    0.12 &    0.16 &    0.56 &    0.23 &  &  \ldots &  \ldots &  \ldots &  \ldots \\
\erii & $-$0.68 &  \ldots &    0.05 &  \ldots &  &  \ldots &  \ldots &  \ldots &  \ldots &  &  \ldots &  \ldots &  \ldots &  \ldots \\
\tmii & $-$1.84 &  \ldots & $-$0.19 &  \ldots &  &  \ldots &  \ldots &  \ldots &  \ldots &  &  \ldots &  \ldots &  \ldots &  \ldots \\
\ybii & $-$0.94 &    0.14 & $-$0.08 &    0.07 &  &  \ldots &  \ldots &  \ldots &  \ldots &  &  \ldots &  \ldots &  \ldots &  \ldots \\
\hfii & $-$0.77 &    0.06 &    0.08 &    0.13 &  & $-$0.27 &    0.11 &    0.42 &    0.00 &  &  \ldots &  \ldots &  \ldots &  \ldots \\
\pbi  &    0.15 &    0.18 &    0.08 &    0.13 &  &    1.40 &    0.21 &    1.18 &    0.32 &  &  \ldots &  \ldots &  \ldots &  \ldots \\
\hline
\end{tabular}
\end{minipage}
\end{table*}

M2 shares similar, and peculiar, characteristics found in the unusual globular
clusters M22, NGC 1851 and $\omega$~Cen, namely, a dispersion in metallicity
and neutron-capture abundance ratios. If a subset of the metal-rich group are
genuine cluster members, then M2 would host stars that span a range in
metallicity from [Fe/H] $\approx$ $-$1.6 to [Fe/H] $\approx$ $-$1.0, a factor of
four.  We note further that even if the most metal-rich stars are not members,
there still remains a metallicity spread of order 0.25 dex among the stars for
which we assert cluster membership. 

M2 appears to be different from M22 and NGC 1851; for the latter two clusters,
the number of stars on the bright subgiant branch is similar to the number on
the faint subgiant branch. In contrast, for M2 the canonical RGB stars
represent the overwhelming majority of stars. As noted in Section 4.4, the
relative numbers of canonical and anomalous RGB stars in M2 is probably
comparable to the relative numbers of bright ($\sim$96\%) and faint ($\sim$4\%)
subgiant branch stars \citep{piotto12}. 

\subsection{Light-, $\alpha$- and Fe-peak elements} 

Regarding the light elements, even with our limited sample it is apparent that
the $r$-only and $r+s$ groups both exhibit star-to-star abundance variations
and correlations between [Na/Fe] and [Al/Fe]. The two populations in M22 and
NGC 1851 both exhibit a O-Na anticorrelation \citep{marino11,carretta11}, and
for $\omega$ Cen, the O-Na anticorrelation is present across a broad
metallicity range \citep{norris95b,johnson10}.  Indeed, every well studied
Galactic globular cluster exhibits star-to-star abundance variations for the
light elements C, N, O, F, Na, Mg and Al (e.g., see reviews by
\citealt{kraft94}; \citealt{gratton04}; \citealt{gratton12}). While these
abundance variations are believed to be produced through hydrogen-burning, the
specific site continues to be debated (e.g.,
\citealt{fenner04,ventura05,decressin07,demink09}). 

For the metal-rich group, the apparent absence of a star-to-star abundance
variation for the light elements is intriguing, although the sample size is
small. No such abundance spread would be expected if these were all field
stars.  On the other hand, a similar situation is present in the
M54+Sagittarius (Sgr) system.  While the O-Na anticorrelation is evident in
M54, the more metal-rich Sgr stars do not exhibit this pattern
\citep{carretta10b}. If the four stars in the metal-rich group are indeed
cluster members, then M2 would share this peculiar feature with M54+Sgr. 

For the $\alpha$ and Fe-peak elements, there is no compelling evidence for a
star-to-star abundance variation within a given group.  Additionally, the
abundance ratios [X/Fe] for a given star are compatible with field stars at the
same metallicity.  In other words, these elements appear to be well-behaved. 

The abundance of Cu offers an important tool to distinguish between field stars
and ``$\omega$ Cen-like'' systems. For M2, the Cu abundance may help establish
additional similarities with $\omega$ Cen and potentially cluster membership,
or otherwise, for the four metal-rich objects for the following reasons. In the
metallicity regime $-$2.0 $\lesssim$ [Fe/H] $\lesssim$ $-$0.5, field stars
exhibit a systematic increase in [Cu/Fe] with increasing metallicity
\citep[e.g.,][]{sneden88,mishenina02,primas08}.  Mono-metallic globular
clusters in the same metallicity range appear to follow the field star trend
\citep{simmerer03}.  $\omega$ Cen, however, displays a near constant Cu
abundance, [Cu/Fe] $\approx$ $-$0.5, over the range $-$1.9 $\le$ [Fe/H] $\le$
$-$0.8 \citep{cunha02}. At higher metallicities, $-$1.2 $\le$ [Fe/H] $\le$
$-$0.4, there is evidence for an increase in the [Cu/Fe] ratio in $\omega$ Cen
\citep{pancino02}, although the rate of that increase is smaller than in field
stars. Chemical evolution models of $\omega$ Cen and the Milky Way by
\citet{romano07} attribute the nucleosynthesis of Cu to massive stars and
successfully reproduce the observed trends. 

\begin{figure*}
\centering
      \includegraphics[width=0.8\hsize]{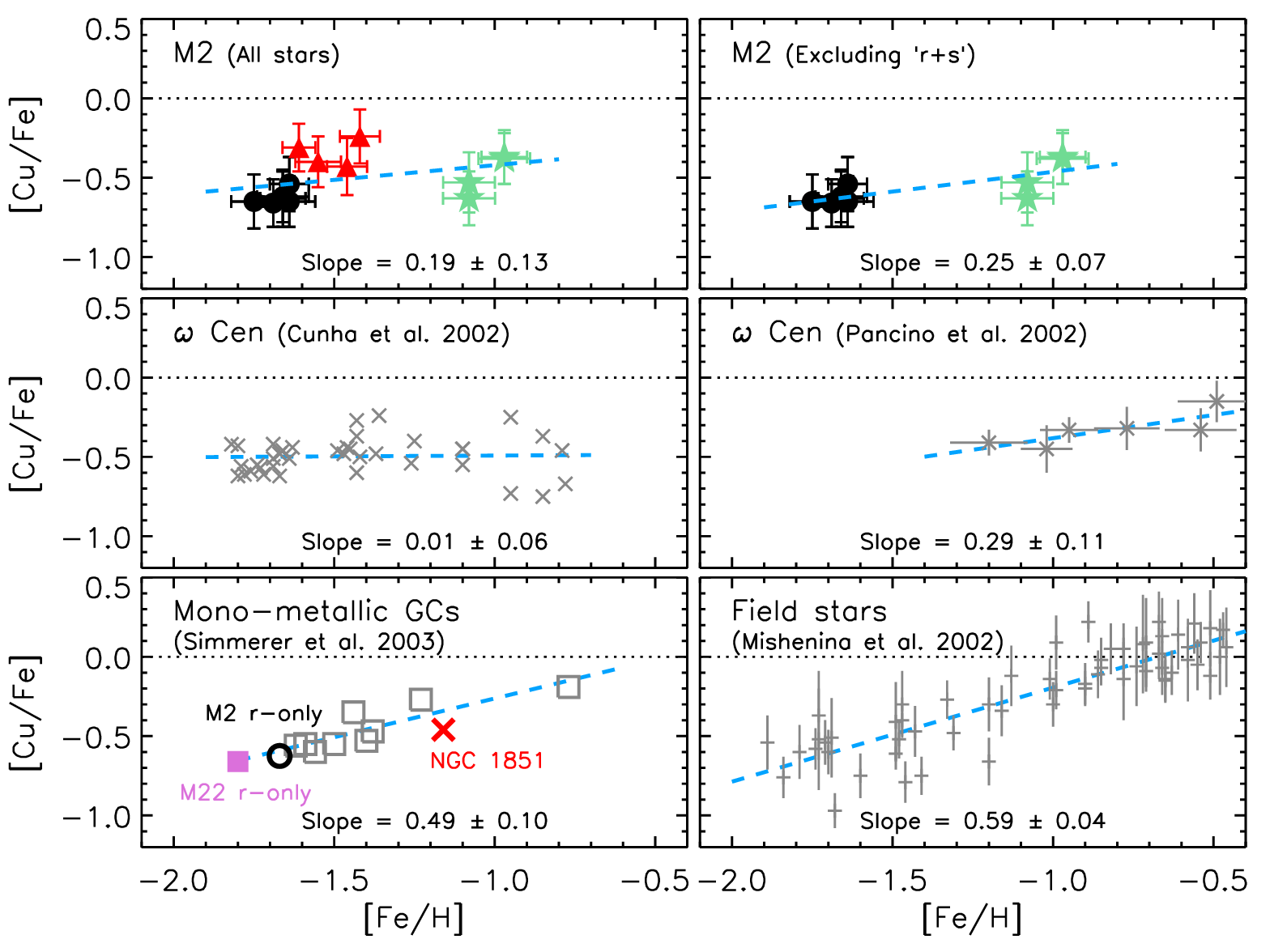}
      \caption{[Cu/Fe] versus [Fe/H] for M2, $\omega$ Cen
\citep{cunha02,pancino02}, mono-metallic globular clusters \citep{simmerer03}
and field stars \citep{mishenina02}. In each panel, we plot the linear fit to
the data and write the slope and uncertainty of the fit. For M2, our values are
adjusted onto the \citet{simmerer03} scale (see text for details). In the lower
left panel, we also include values for the M2 $r$-only (metal-poor) group, M22
$r$-only group \citep{roederer11} and NGC 1851 \citep{carretta11}, although
these values are not included in the linear fit to the data. 
      \label{fig:cu} }
\end{figure*}

In Figure \ref{fig:cu}, we plot [Cu/Fe] versus [Fe/H] for M2, field stars
\citep{mishenina02}, mono-metallic globular clusters \citep{simmerer03} and
$\omega$ Cen \citep{cunha02,pancino02}. The Cu abundances in M2 displayed in
this figure have been adjusted in the following manner. Following
\citet{simmerer03}, the abundances from the 5105\,\AA\ and 5782\,\AA\ lines are
referenced to solar values of $\log \epsilon_\odot$ = 4.21 and $\log
\epsilon_\odot$ = 4.06, respectively. Such an approach reflects the different
solar abundances obtained from these lines, and we note that the abundances we
derive for program stars from the 5782\,\AA\ line are, on average, 0.21 dex
$\pm$ 0.03 dex ($\sigma$ = 0.11 dex) higher than those from the 5105\,\AA\
line. The gradient of the linear fit to M2 is not affected by these zero-point
offsets. We also stress that although this figure includes data from numerous
studies, the linear fit in each panel is performed upon data obtained from a
single study (for the lower left panel, M2, M22 and NGC 1851 are from different
studies but those data are not included in the linear fit).  So long as each
sample is analysed uniformly, the slopes should be robust and we can compare
them in a quantitative manner.  In Figure \ref{fig:cu}, the slopes for the
field stars and mono-metallic globular clusters are in good agreement, and
these slopes differ from that seen in $\omega$ Cen. The behaviour of the slope
of [Cu/Fe] versus [Fe/H] in M2---whether or not the $r+s$ stars are
considered---is different from the field stars, mono-metallic globular clusters
and $\omega$ Cen over the metallicity range $-$1.7 $\leq$ [Fe/H] $\leq$ $-0.9$.
The metal-rich stars, relative to the $r$-only and $r+s$ groups, do not follow
the field star trend, and this may be the strongest abundance-based evidence
that they are cluster members.  Furthermore, if the metal-rich stars are indeed
cluster members, then M2 does not share a similar chemical enrichment history
to $\omega$ Cen, at least for Cu.  Figure \ref{fig:cu} also demonstrates that
the mean [Cu/Fe] ratios in NGC 1851 \citep{carretta11} and the $r$-only groups
in M2 and M22 \citep{roederer11} match the trends established by the
mono-metallic globular clusters. The Cu in the $r+s$ group of stars may include
small contributions from $s$-process nucleosynthesis, so we do not discuss Cu
in the $r+s$ group here. 

\subsection{Neutron-capture abundance patterns in M2}
\label{ncap}

We now turn our attention to the neutron-capture elements. In the abundance
analysis described in Section 3.2, we examined up to 122~lines of elements with
atomic numbers $Z \geq$~38 in each of the program stars. All abundances were 
computed by matching synthetic spectra, generated using one dimensional
plane-parallel model atmospheres, to the observed spectra under the assumption
that LTE holds in the line-forming layers. 

The abundances of Sr and
Pb were derived from neutral lines. For the program stars, Sr~\textsc{i} and
Pb~\textsc{i} are minority species, and LTE calculations will tend to
underestimate the populations of the lower levels of the Sr~\textsc{i}
4607\,\AA\ and Pb~\textsc{i} 4057\,\AA\ transitions.  Abundances of strontium
and lead derived in LTE from these lines are thus underestimated.  Calculations
allowing for departures from LTE in the line-forming layers by making
reasonable assumptions for the photoionization cross sections have been made
for stars with stellar parameters similar to those in our sample.  These
non-LTE calculations suggest that our LTE analysis may underestimate the
strontium abundance by $\approx$~0.3$-$0.5~dex \citep{bergemann12,hansen13} and
the lead abundance by $\approx$~0.3$-$0.4~dex \citep{mashonkina12}. 
The values presented in our tables and figures reflect the LTE values.
Neglecting the non-LTE corrections for these two elements should not
significantly affect any of the abundance differences between the $r$-only and
$r+s$ groups of stars that we shall discuss below. 

\begin{figure*}
\centering
      \includegraphics[width=0.95\hsize]{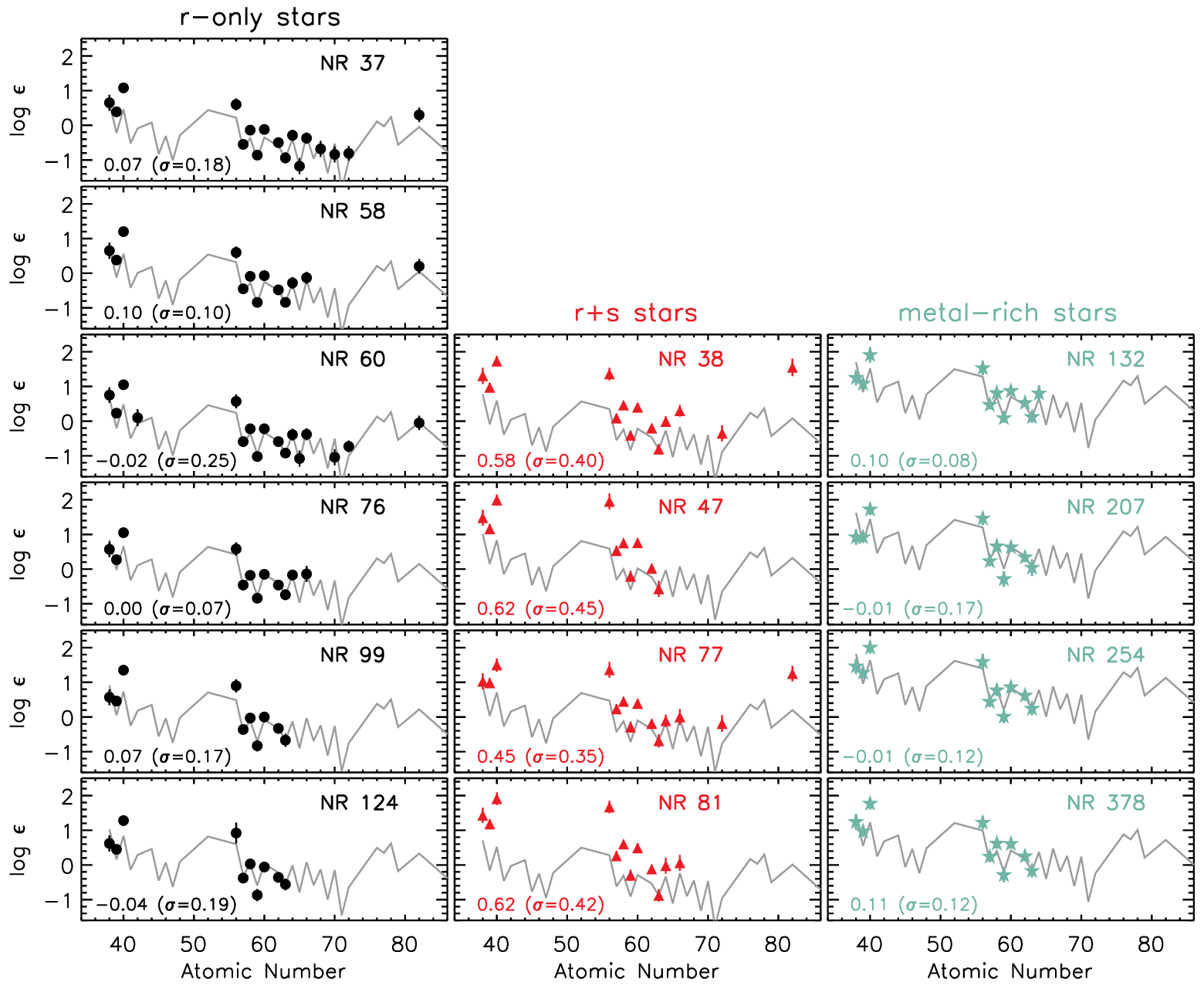}
      \caption{Logarithmic abundances for $Z \geq$~38 elements in the six 
$r$-only stars (black circles, left panels), the four $r+s$ stars (red
triangles, middle panels) and the four metal-rich stars (aqua stars, right
panels). For zirconium, the abundance plotted is the value derived from
Zr~\textsc{ii} lines.  The grey line illustrates the abundances in the \rpro\
standard star \bd\ \citep{cowan02,cowan05,sneden09,roederer10a,roederer12a}
normalised to the europium abundance in each star.  Lead has not been detected
in \bd, so we instead show the predicted Pb/Eu ratio based on the average Pb/Eu
observed in Figure~3 of \citet{roederer10b}. The numbers in the lower left
corner of each panel are the mean difference ``star $-$ \bd'' and the
dispersion (standard deviation) for the elements from Ba to Hf. 
      \label{starplot} }
\end{figure*}

Figure~\ref{starplot} illustrates the heavy element abundance patterns found in
each star of our sample. The six stars shown in the left-hand panels are those
on the canonical RGB, the four stars in the middle panels are the
neutron-capture rich anomalous RGB stars and the four stars in the right-hand
panels are the metal-rich anomalous RGB stars. For comparison, in each panel of
this figure we overplot the heavy element abundance pattern found in the \rpro\
rich standard star \bd\ (normalised to the Eu abundance), whose metallicity is
only a factor of $\approx$~2.5 lower than the majority of stars in M2. The
stars on the canonical RGB have heavy element abundance patterns very similar
to one another and to the \rpro\ pattern in \bd, and the overall amounts of
heavy elements are constant within their mutual uncertainties.  We refer to
these six stars (NR~37, NR~58, NR~60, NR~76, NR~99 and NR~124) on the canonical
RGB as the ``$r$-only group.'' The reasoning behind this name will be made
clear shortly. 

As shown in the middle panels of Figure~\ref{starplot}, the heavy elements
in the neutron-capture rich anomalous RGB stars in M2 tell a different story.  
All heavy elements except europium in these four stars exhibit noticeable
abundance enhancements relative to the stars on the canonical RGB and,
therefore, enhancements relative to the \rpro\ standard \bd. The pattern
changes little from one star to the next, and the overall abundances in this
group of stars are also constant within their mutual uncertainties.  We refer
to these four stars (NR 38, NR 47, NR 77 and NR 81) as the ``$r+s$ group.'' 

\begin{figure}
\centering
	\includegraphics[width=0.99\hsize]{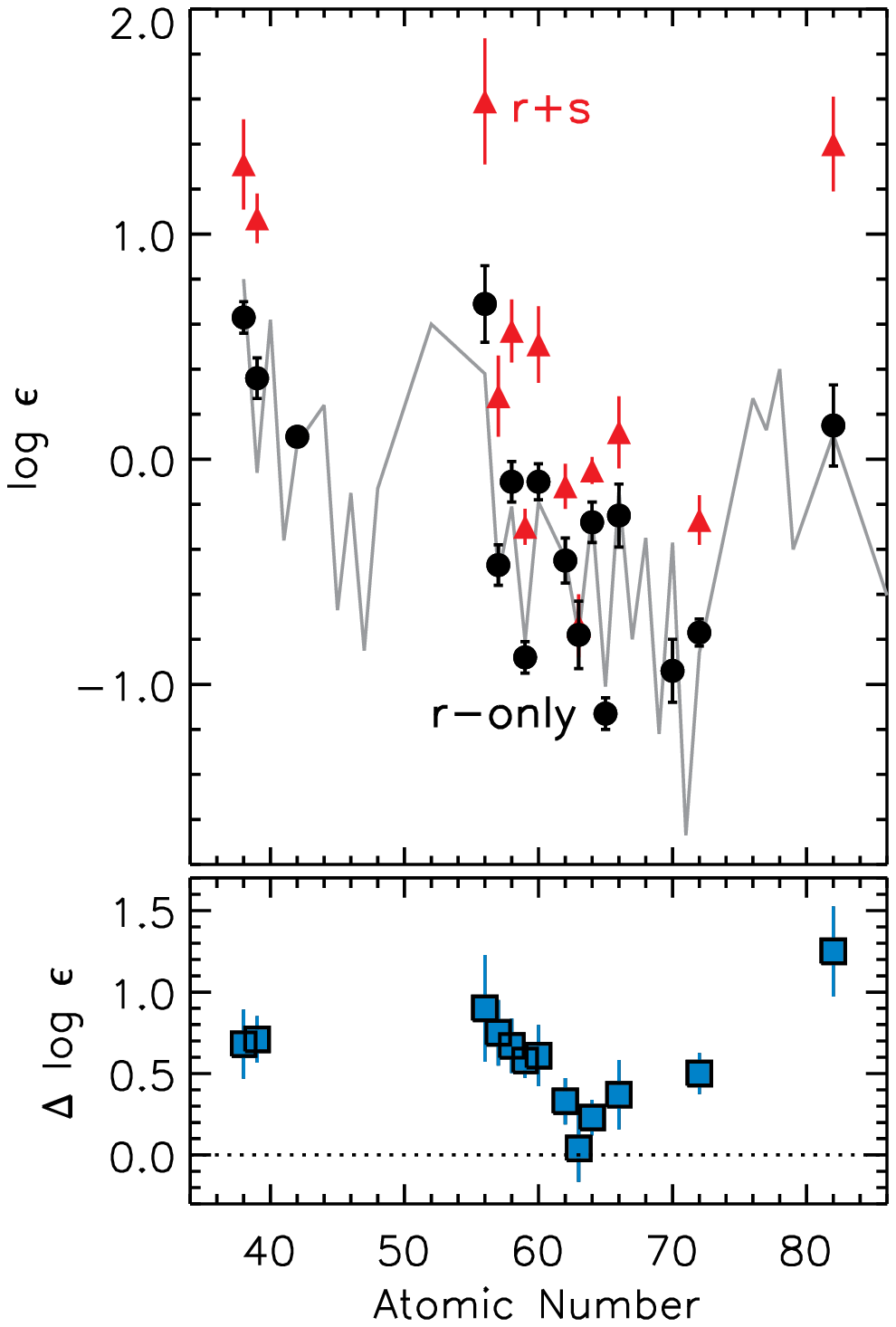}
      \caption{Top panel: Mean logarithmic abundances for the six $r$-only
stars (black circles) and the four $r+s$ stars (red triangles).  (Only elements
measured in more than one star are included, i.e., we exclude Mo, Er and Tm.)
The zirconium abundance derived from Zr~\textsc{ii} lines is shown.  The gray
line illustrates the abundances in the \rpro\ standard star \bd\ normalized to
the europium abundance.  Bottom panel: Differences in these mean abundances.
The dotted line indicates zero difference. 
      \label{meanplot} }
\end{figure}

The consistent patterns and levels of enhancement found within each of the
$r$-only and $r+s$ groups suggests that we can average together their
abundances to reduce the random uncertainties, which is especially helpful for
elements whose abundances are derived from small numbers of lines.  These mean
abundance patterns are listed in Table~\ref{tab:meanabun} and illustrated in
Figure~\ref{meanplot}.  Subtle differences between the stars in the $r$-only
group and \bd, (e.g., small overabundances in M2 for strontium, yttrium,
zirconium, barium, cerium and neodymium, as well as small underabundances in M2
for ytterbium) may simply reflect differing combinations of material produced
by the so-called weak and main components of the \rpro\ enriching M2 and \bd.
This is plausible because the overall level of $r$-process enhancement relative
to iron is different in \bd\ and M2, with [Eu/Fe] = +0.9 and +0.4,
respectively. Regardless, Figure~\ref{starplot} demonstrates that the heavy
elements in the stars in the $r$-only group in M2 owe their origin to \rpro\
nucleosynthesis with little or no \spro\ contributions. 

\begin{figure*}
\centering
	\includegraphics[width=0.9\hsize]{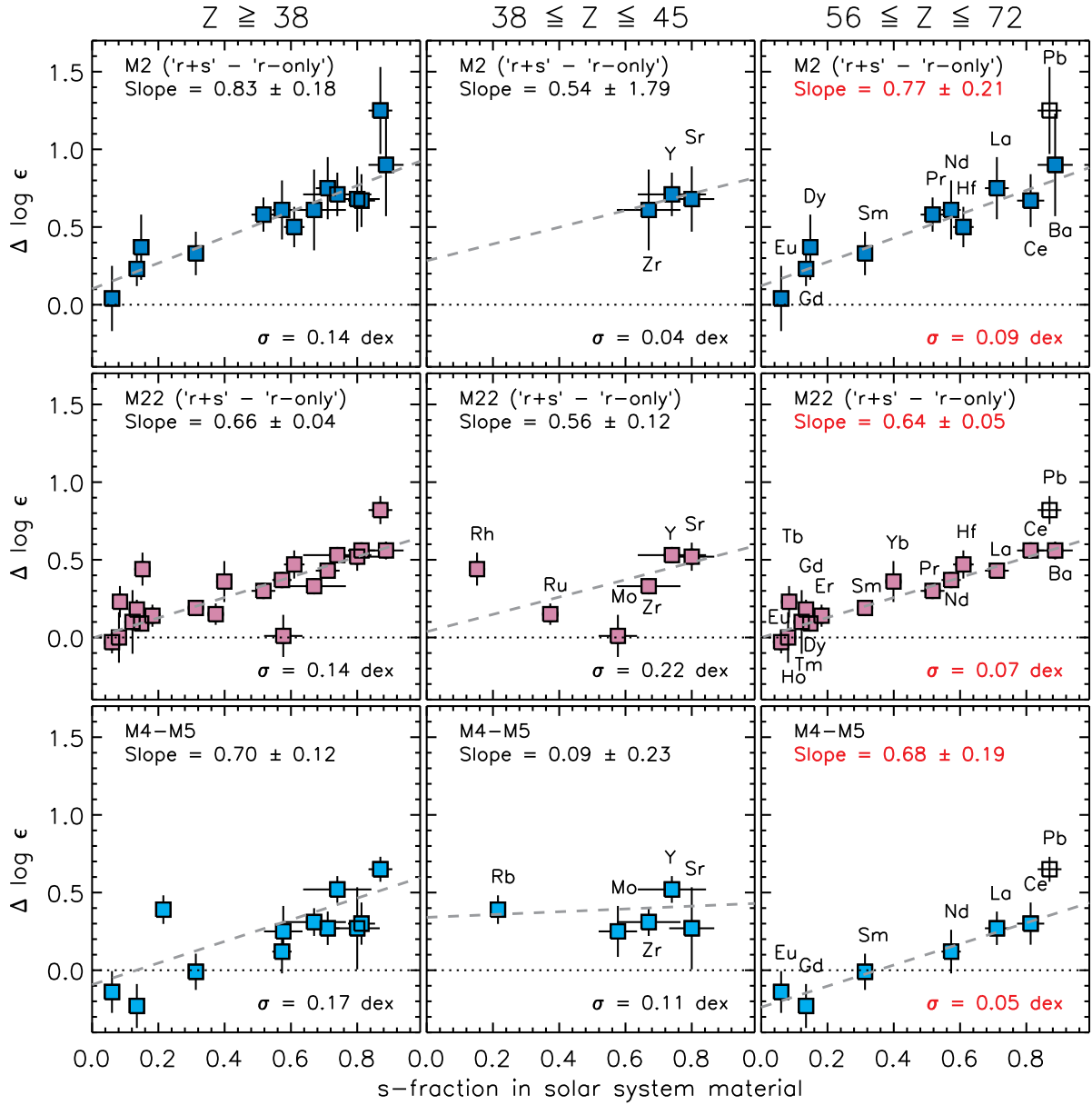}
      \caption{Differences in the mean abundances between the $r+s$ group and
the $r$-only group as a function of the fraction of each element attributed to
an \spro\ origin in solar material \citep{bisterzo11}. M2, M22 and M4$-$M5 are
shown in the upper, middle and lower panels, respectively. The elements $Z$
$\geq$ 38, 38 $\le$ $Z$ $\le$ 45 and 56 $\le$ $Z$ $\le$ 72 are displayed in the
left-hand, middle and right-hand panels, respectively. The dotted line
indicates zero difference. In each panel we overplot the linear fit to the data
and write the slope and error as well as the dispersion about the fit. In the
right panels, we plot Pb (Z = 82), although this element is not
included in the fit.  
\label{sfracplot} } 
\end{figure*}

In contrast, the middle panels of Figure \ref{starplot} demonstrate that the
$r+s$ stars have abundance patterns that are inconsistent with \rpro\
nucleosynthesis alone. Figure~\ref{sfracplot} demonstrates that the excess of
heavy elements ($Z \geq$~38) found in the $r+s$ group relative to the $r$-only
group exhibits an unmistakable correlation with the fraction of each element
attributed to an \spro\ origin in solar system material.  Elements with a high
\spro\ fraction in the solar system are most overabundant in the $r+s$ group in
M2, and those with small \spro\ fractions in the solar system show little
excess.  Only a few percent of the europium in the solar system is attributed
to \spro\ nucleosynthesis, and this element shows a constant abundance in both
the $r$-only and $r+s$ groups of stars; the average $\log\epsilon$ (Eu) and
[Eu/Fe] ratios are very similar between these two groups.  This suggests that
there is a common \rpro\ abundance foundation in each star in the $r$-only and
$r+s$ groups in M2. (We set aside, for now, the metal-rich group.) Following
\citet{roederer11}, we speculate that the stars in the $r+s$ group formed from
additional material enriched by products of \spro\ nucleosynthesis as well as
iron.  We emphasise that the abundance differences between these two groups of
stars should be nearly insensitive to any non-LTE effects given the modest
range in stellar parameters (\teff, \logg, [Fe/H]) spanned by the sample. 

We now consider the metal-rich group (stars NR 132, NR 207, NR 254 and NR 378)
shown in the right-hand panels of Figure \ref{starplot} noting that these stars
may, or may not, be cluster members. The heavy element abundances in the
metal-rich group closely resemble an \rpro\ pattern, despite the fact that the
overall metallicity of these objects is a factor of $\approx$~4 higher than the
other cluster members.  Furthermore, the [X/Fe] ratios (where X denotes any
element with $Z \geq$~56) in the metal-rich group are on average 0.17 dex $\pm$
0.02 dex ($\sigma$ = 0.05 dex) higher than in the $r$-only stars, i.e., a
factor of $\approx$1.5.  If we assume that the overall metal content in an
isolated stellar system increases monotonically with time, the metal-rich group
should have formed later than either the $r$-only or $r+s$ groups. To the best
of our knowledge, no isolated self-enriched stellar system shows a return to
\rpro\ dominance after previous enrichment by a substantial amount of \spro\
material. We conclude that the four metal-rich stars cannot be easily
understood as members of a single self-enriched stellar system.  That said, the
data do not preclude a scenario in which M2 is composed of independent
fragments that experienced different chemical enrichment histories
\citep{searle77}. 

\subsection{The origin of the $s$-process material}
\label{sproorigin}

If we assume that the \rpro\ enrichment is common to both the $r$-only and
$r+s$ groups in M2, we can subtract the average abundances found in the
$r$-only group from those in the $r+s$ group to obtain the intrinsic abundance
ratios of the \spro\ material added to the $r+s$ group.  These differences are
shown in Figure \ref{meanplot} (as a function of atomic number) and in Figure 
\ref{sfracplot} (as a function of the fraction of each element attributed to
the $s$-process in solar system material). 

\citet{roederer11} performed a similar calculation for the two stellar groups
in M22 and the unrelated clusters M4 and M5 using data from
\citet{yong08m4m5a,yong08m4m5b}. In Figure \ref{sfracplot}, we include the
abundance differences for the M22 groups as well as the abundance
differences when subtracting the mean values for M5 from those of M4 which we
denote as ``M4 $-$ M5''\footnote{These are two well-studied unrelated
clusters of similar metallicity, [Fe/H] $\simeq$ $-$1.2, and M4 is known to
exhibit a moderate enhancement in $s$-process element abundances compared to M5
\citep{ivans99,ivans01}. As in \citet{roederer11}, subtracting the abundances
for M5 from M4 attempts to quantify the $s$-process contribution to M4.}. 
In this
figure we adopt the values from \citet{bisterzo11} for the fraction of each
element attributed to the $s$-process in solar system material. As noted, there
is a clear trend between the abundance differences and the $s$-process
fraction. As originally proposed by \citet{roederer11}, we argue that the
abundance residual represents $s$-process material. 

When considering all elements with $Z$ $\ge$ 38 (i.e., Sr and heavier
elements), the three ``systems,'' (1) $\langle r+s \rangle$ $-$ $\langle
r$-only$\rangle$ in M2, (2) $\langle r+s \rangle$ $-$ $\langle r$-only$\rangle$
in M22 and (3) $\langle$M4$\rangle$ $-$ $\langle$M5$\rangle$, exhibit identical
gradients within their mutual uncertainties. Such a result is surprising given
that the yields for the $s$-process elements in AGB stars are mass and
metallicity dependent (e.g., \citealt{busso01,cristallo11,karakas12}; Fishlock
et al., in preparation). If our interpretation that the abundance
residuals in these systems represent $s$-process material is correct, then the
implication is that these three systems, which span a range in metallicity from
[Fe/H] $\approx$ $-$1.8 to [Fe/H] $\approx$ $-$1.2, experienced enrichment by
$s$-process material of indistinguishable composition. Quantitative chemical
evolution modelling is needed to test this intriguing hypothesis, and
Shingles et al.\ (in preparation) are investigating M22 and M4 and comparing
the predicted and observed enrichment patterns taking into account yields from
AGB and massive stars. 

If we consider only elements with 38 $\le$ $Z$ $\le$ 45 (i.e., Sr to Rh), the
gradients do not exhibit any consistent patterns.  In sharp contrast, however,
the elements from 56 $\le$ $Z$ $\le$ 72 (i.e., Ba to Hf) exhibit identical
gradients within their mutual uncertainties. For these elements, the measured
abundance differences in each system are consistent with a single relation. 
This implies that the enrichment in M2, M22 and M4 involved $s$-process
material of remarkably similar composition despite the factor of $\sim$~4
difference in metallicity. 

The intrinsic \spro\ ratios and indices\footnote{We adopt the indices as
defined by \citet{bisterzo10}: the ratios of light ($ls$) and heavy ($hs$)
\spro\ abundances are [$ls$/Fe]~$\equiv \frac{1}{2}$([Y/Fe]~$+$~[Zr/Fe]) and
[$hs$/Fe]~$\equiv \frac{1}{3}$([La/Fe]~$+$~[Nd/Fe]~$+$~[Sm/Fe]).  These include
elements at the first (Sr, Y, Zr) and second (Ba, La, Ce, Pr, Nd) \spro\ peaks;
Pb is the sole representative of the third \spro\ peak.  Similarly,
[$hs$/$ls$]~$\equiv$~[$hs$/Fe]~$-$~[$ls$/Fe].} are 
[Pb/La]$_{\rm s}$~$= 	 +$0.53 in M2,
			$+$0.18 in M22 and
			$-$0.01 in M4 $-$ M5;
[$hs$/$ls$]$_{\rm s}$~$=-$ 0.02 in M2,
			$-$0.01 in M22 and
			$-$0.50 in M4 $-$ M5; and
[Pb/$hs$]$_{\rm s}$~$=   +$0.72 in M2,
			$+$0.29 in M22 and
			$+$0.28 in M4 $-$ M5.
Uncertainties on these ratios are typically 0.1$-$0.2~dex.  These ratios and
indices are largely insensitive to uncertainties in the atomic data and non-LTE
effects.  For [Pb/La], M2 exhibits a higher ratio than M22 and M4 $-$ M5. For
[$hs$/$ls$], M4 $-$ M5 exhibits lower ratios than M22 and M2, although this may
reflect the higher metallicity of M4 and M5 relative to the other two clusters.
For [Pb/$hs$], M2 exhibits a higher ratio than the other two systems. 

A number of studies have investigated \spro\ nucleosynthesis in metal-poor
stars on the AGB \citep[][Fishlock et al., in
preparation]{goriely00,goriely01,cristallo09,cristallo11,bisterzo10}. While
most of these models fail to offer an exact match for the metallicity of M2, we
can use them to get a sense of \spro\ nucleosynthesis ratios predicted for
metallicities higher and lower than M2. We find encouraging agreement when
comparing our results to the [$hs$/$ls$] and [Pb/$hs$] indices presented in
Figures~C3 and C5 of \citet{bisterzo10} for 3 and 5~\msun\ AGB stars at the
appropriate metallicities for M2, M22 and M4 $-$ M5. Furthermore, we note that
the yields of Fishlock et al.\ (in preparation) for their 3~\msun\ and
3.5~\msun\ models for [Fe/H] = $-$1.2 bracket the [Pb/La]$_{\rm s}$,
[$hs$/$ls$]$_{\rm s}$ and [Pb/$hs$]$_{\rm s}$ ratios in M2, M22 and M4 $-$ M5.
Quantitative chemical evolution models based on their yields would be of great
interest.  Overall, we reach the same conclusion drawn by \citet{roederer11}:
AGB stars with masses less than 3~\msun\ cannot reproduce the observed ratios
unless the standard $^{13}$C pocket efficiency in the models is reduced by
factors of 30 or more. 

If we assume that the stars in the $r+s$ group in M2 formed later than the
stars in the $r$-only group, and that the AGB stars responsible for
distributing this \spro\ material in M2 formed simultaneously with the stars in
the $r$-only group, this sets an upper limit on the amount of time that passed
between the formation of the $r$-only group and the $r+s$ group.  For a
3~\msun\ AGB star, adopting the approximate stellar lifetimes computed by
\citet{mowlavi12}, this sets a limit of no more than 300~Myr or so between the
two groups.  Of course, this limit would be even smaller if higher-mass AGB
stars were the source of the \spro\ material. Finally, although we have
focused on the neutron-capture elements, the difference in [Fe/H] between the
$r$-only and $r+s$ groups requires some source(s) that produces the elements
from Si to Zn (and perhaps other elements) to increase the abundances of these
elements between these groups.  

\section{CONCLUSION}

In this paper we present a spectroscopic analysis of giant stars in the
multiple population globular cluster M2. Our principal and novel results
include the following. First, we identify a star-to-star dispersion in iron
abundance with the anomalous RGB stars (i.e., stars lying redward of the
dominant RGB) being more metal-rich than the canonical RGB objects. The iron
abundance distribution has a dominant peak at [Fe/H] $\approx$ $-$1.7 and
smaller peaks at $-$1.5 and $-$1.0, although membership for the latter group
remains to be established.  Secondly, the neutron-capture element abundances
exhibit a star-to-star dispersion with a possible bimodal distribution. In this
regard, M2 is chemically similar to the globular clusters M22, NGC 1851 and
$\omega$ Cen, whose subgiant branches exhibit multiple sequences. It is likely
that M2 has therefore experienced a similarly complex formation history.
Thirdly, when subtracting the average abundances in the $r$-only group from
those of the $r+s$ group, the abundance residual exhibits a striking
correlation with the fraction of each element attributed to the $s$-process in
solar system material. This residual is remarkably similar to that found in M22
and in M4 $-$ M5. Such a similarity would indicate that M2, M22, and M4 were
enriched by $s$-process material of identical composition and potentially
offers important observational constraints on the nature of the $s$-process in
low metallicity environments. A comparison with theoretical predictions reveals
that AGB stars with masses less than 3 \msun\ are unlikely to have played a
major role in the chemical enrichment of M2. In addition to the AGB star
contribution, some source(s) is needed to increase the abundances of the
elements from Si to Zn in the $r+s$ group relative to the $r$-only group.
Additional studies are essential to understand the formation and evolution of
this complex cluster. 

\section*{Acknowledgments}

We thank K.\ Cudworth for providing electronic data and the referee, Kim
Venn, for helpful comments. 
Some of the AAT data were taken in time allocated to the AEGIS program (PI
Keller).  The authors are grateful to the AEGIS team for access to these data. 
D.Y, G.D.C, A.I.K, J.E.N, A.F.M and A.P.M gratefully acknowledge support from
the Australian Research Council (grants DP0984924, FT110100475, DP120100475,
DP120100991 and DP120101237). 
Funding for the Stellar Astrophysics Centre is provided by The Danish National
Research Foundation. The research is supported by the ASTERISK project
(ASTERoseismic Investigations with SONG and Kepler) funded by the European
Research Council (Grant agreement no.: 267864).

\label{lastpage}

\end{document}